\documentclass[reprint,aps,prc,amsmath,amssymb,showpacs,showkeys,twocolumns,floatfix,diagbox]{revtex4-1}
\usepackage{graphicx}
\usepackage{dcolumn}
\usepackage{bm}
\usepackage{longtable}
\usepackage[flushleft]{threeparttable}
\usepackage{hyperref}
\usepackage{siunitx}
\usepackage{float}
\begin{document}

\preprint{APS/123-QED}

\title{Manifestation of deformation and nonlocality in $\alpha$ and cluster decay}

\author{D. F. Rojas-Gamboa}
\email{df.rojas11@uniandes.edu.co}
\affiliation{Departamento de F\'isica, Universidad de Los Andes,
Carrera 1 No.18A-10, Bogot\'a 111711, Colombia}
\author{J. E. Perez Velasquez}
\email{je.perez43@uniandes.edu.co}
\affiliation{Departamento de F\'isica, Universidad de Los Andes,
Carrera 1 No.18A-10, Bogot\'a 111711, Colombia}
\author{N. G. Kelkar}
\email{nkelkar@uniandes.edu.co}
\affiliation{Departamento de F\'isica, Universidad de Los Andes,
Carrera 1 No.18A-10, Bogot\'a 111711, Colombia}
\author{N. J. Upadhyay}
\email{nupadhyay@mum.amity.edu}
\affiliation{Department of Sciences, Amrita School of Engineering, Amrita Vishwa Vidyapeetham, Chennai 601103, India}

\begin{abstract}
It is common to study the strong decay of a heavy nucleus as a tunneling phenomenon where the $\alpha$ ($^4$He) or a light nuclear cluster tunnels through the Coulomb barrier formed by its interaction with the heavier daughter nucleus. The position and width of the Coulomb barrier is determined by the total interaction potential between the two daughter nuclei. We examine the effects of including nonlocality and deformation in the interaction potential by calculating the half-lives, $t_{1/2}$, and thereby phenomenological preformation factors of several nuclei which have the possibility of decaying by emitting either an $\alpha$ or a light nuclear cluster. The effect of deformation manifests itself by a decrease in $t_{1/2}$ for all the decays studied. The effect of nonlocality is studied within two different models of the nuclear potential: the energy independent but angular momentum ($l$) dependent Mumbai (M) potential and the energy dependent Perey-Buck (PB) potential. The nonlocal nuclear interaction leads to a decrease in all half-lives studied. Though the decrease is larger due to the Perey-Buck potential, half-lives evaluated using the Mumbai potential show a strong sensitivity to the $l$ value in the decay. This feature of the $\alpha$ and cluster decay half-lives can provide a complementary tool in addition to scattering data which are more commonly used to fix the parameters of a nonlocal potential in literature. 
\end{abstract}

\pacs{23.60.+e, 21.10.Tg, 21.30.-x}

\keywords{}

\maketitle

\section{Introduction}\label{intro}
$\alpha$ decay of radioactive nuclei, discovered in 1899 by Rutherford \cite{ruth}, is one of the most important decay modes in providing information about the structure of heavy nuclei such as the radius, deformation, and shell effects \cite{NiRenPRC80,QianRen,PMohr}. In addition to this, cluster radioactivity (CR) has also been very instrumental in providing structure information of heavy nuclei \cite{seif2021}. The spontaneous emission of light fragments heavier than the $\alpha$ particle, but lighter than a typical mass of a light fragment in the fission process (A $\gtrsim$ 60) is referred to as cluster radioactivity or cluster decay. Theoretically predicted cluster decay \cite{Sandulescu1980} was experimentally confirmed in 1984 by Rose and Jones \cite{Rose1984}. Study of both the $\alpha$ and cluster decays for the isotopes of the same element provides insight about the impact of shell effects and deformations on the probabilities of these processes. In an investigation involving the comparison of different models for cluster radioactivity in superheavy nuclei \cite{zhang}, the authors found that a universal decay law which can predict light and heavy CR leads to the shortest half-lives when the daughter nuclei are around the doubly magic $^{208}$Pb nucleus. Another interesting conclusion was that the CR dominates over $\alpha$ decay for $Z \ge$ 118 nuclei.  Because of the large difference in the masses of the cluster and the daughter nucleus, just like the $\alpha$ decay, the cluster decay is also treated as a tunneling problem for the calculation of half-lives of nuclei.

In a recent work \cite{Johan2019}, we have investigated the effects of nonlocality on $\alpha$-decay half-lives of nuclei. It is observed that the effective potential obtained between the $\alpha$ and the daughter nucleus within the nonlocal framework decreases the half-lives. This study was limited to $\alpha$ decay involving spherical nuclei and deformation was not considered. In the present work, we study the nonlocal effect for the $\alpha$ and cluster decay modes of a given parent nucleus. Furthermore, we also study the impact of nonlocality in the presence of deformation which modifies the barrier characteristics. In the $\alpha$ decay, the shape of the daughter nucleus contributes to the deformation term, whereas in the cluster decay, the shape of the lighter cluster in the decay channel contributes to the deformation.

The article is organized as follows. In Sec. \ref{formalism}, we give a brief overview of the density-dependent double-folding (DF) model used to evaluate the potential between the cluster and the daughter nucleus followed by the forma\-lism for the evaluation of decay half-lives. The two mo\-dels used to study the nonlocal effects in the $\alpha$ and cluster decay of some heavy nuclei are also discussed. In order to extend the half-life calculations in the pre\-sence of deformation, we provide the details about the deformed DF nuclear potential and calculation of the half-lives from the deformed decay width. In Sec. \ref{results}, we present the results and discuss them. Finally, in Sec. \ref{summar} we summarize our findings.

\section{Formalism for $\alpha$ and cluster decay} \label{formalism}
In previous studies on $\alpha$ and cluster decay \cite{NiRenPRC80,QianRen,PMohr}, it has been found that the half-life is an important observable related to the nuclear structure. In most of the theoretical models, the $\alpha$ and cluster decay half-life of a nucleus is evaluated as
\begin{equation}\label{eq:half-life}
    t_{1/2}=\frac{\hbar\ln 2}{\Gamma}\,,
\end{equation}
where the decay width $\Gamma$ can be calculated within the framework of the semiclassical Jeffreys-Wentzel–Kramers–Brillouin (JWKB) approximation \cite{Gurvitz1987,Castaneda2007, Kelkar2016}, as
\begin{equation}\label{eq:Gamma}
    \Gamma=P_{i}\,\frac{\hbar^{2}}{2 \mu}\left[\int_{r_{1}}^{r_{2}} \frac{{\rm d} r}{k(r)}\right]^{-1}\exp\left[-2 \int_{r_{2}}^{r_{3}} k(r)\,{\rm d} r\right]\,,
\end{equation}
where $k(r)=\sqrt{\frac{2 \mu}{\hbar^{2}}|V(r)-E|}$. $P_i$ (with $i$ = $\alpha$ or $c$) is the preformation factor in $\alpha$ or cluster ($c$) decay, $\mu$ is the reduced mass of the cluster-daughter system and $r_i$ ($i = 1, 2, 3$) are the three classical turning points. These latter are solutions of $V(r_i)=E$, where the energy $E$ of the $\alpha$ particle or cluster is usually taken to be the same as the $Q$-value for the decay. The exponential factor $\exp{[-2 \int_{r_2}^{r_3} k(r)\, {\rm d}r]}$ is known as the penetration probability and the factor $[\int_{r_{1}}^{r_{2}} {\rm d}r / k(r)]^{-1}$ in front of this arises from the normalization of the bound-state wave function in the region between the turning points $r_1$ and $r_2$ \cite{Gurvitz1987,Castaneda2007,Kelkar2016}. An essential ingredient for evaluating both factors is the interaction potential $V(r)$ between the emitted particle and the daughter nucleus that will be explained below. The appearance of a preformation factor $P_i$ ($i$ = $\alpha$ or $c$) in Eq. (\ref{eq:Gamma}) takes into account the non-zero probability for the existence of the preformed cluster or $\alpha$ particle inside the parent as suggested by several theoretical models \cite{Xu-etal2016,Xu-etal2017}. In the present study, we compare the effects of nonlocalities and deformation in different mo\-dels. Hence, we shall set $P_i = 1$ when we compare the half-lives calculated within the different models.

\subsection{Double folding model}\label{DFpot}
\begin{figure}[h!]
    \centering
    \includegraphics{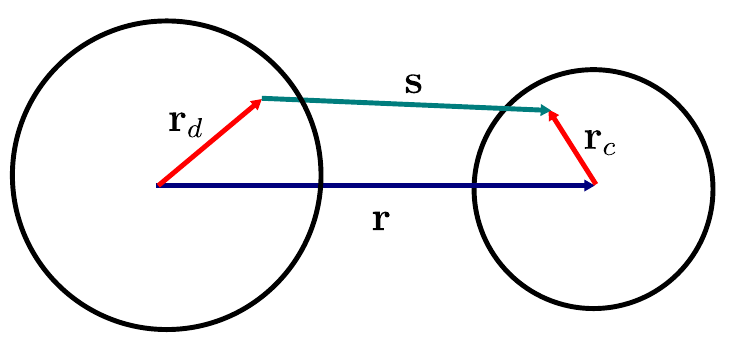}
    \caption{Coordinates used in the DF calculations. Vectors $\mathbf{r}_{c}$ and $\mathbf{r}_{d}$ correspond to the position of a nucleon in the cluster and daughter nucleus, respectively.}
    \label{fig:spherical_nuclei}
\end{figure}

$\alpha$ and cluster decay are treated as a tunneling problem where the emitted particle, preformed inside the parent nucleus, tunnels through the potential barrier. The interaction potential between the daughter nucleus and the cluster consists of the nuclear $V_N$, Coulomb $V_C$, and centrifugal potentials and is given by
\begin{equation}\label{eq:totalpotential}
    V(r)=\lambda\, V_{N}(r)+V_{C}(r)+\frac{\hbar^{2}\left(l+\frac{1}{2}\right)^{2}}{2\mu r^{2}}\,,
\end{equation}
where $r$ is the separation between the center of mass of the cluster and the center of mass of the daughter nucleus, $l$ is the angular momentum carried by the emitted cluster, and $\lambda$ is the strength of the nuclear interaction (see Fig. \ref{fig:spherical_nuclei}). The latter has been left as a parameter in order to ensure that the Bohr-Sommerfeld quantization condition \cite{Castaneda2007} is satisfied. Thus $\lambda$ fixes the strength of the nuclear potential. The centrifugal term containing, $l(l+1)$, is replaced by the correction, $(l+\frac{1}{2})^2$, introduced by Langer when one uses the semiclassical approximation \cite{Langer1937}. This is essential to ensure the correct behavior of the JWKB wave function near the origin. The potential used in the calculation then has three turning points, $r_1$, $r_2$, and $r_3$.

For the nuclear potential it is desirable to relate the nucleus-nucleus interaction to the NN interaction. This can be obtained by folding an effective NN interaction over the density distribution of the two nuclei. This model is known as the double folding (DF) model \cite{Satchler1979}. It has been used widely and has been found to be quite successful in evaluating the $\alpha$ decay half-lives \cite{Castaneda2007, Kelkar2016, Johan2019}. The folded nuclear potential is written as
\begin{equation}\label{eq:V_Double-Folding}
    V_N(\mathbf{r})=\int \mathrm{d} \mathbf{r}_{c} \mathrm{d} \mathbf{r}_{d} \,\rho_{1}\left(\mathbf{r}_{c}\right)v_N\left(|\mathbf{s}|=\left|\mathbf{r}+ \mathbf{r}_{c}-\mathbf{r}_{d}\right|\right) \rho_{2}\left(\mathbf{r}_{d}\right)\,.
\end{equation}
where $\rho_i$ ($i=$ 1, 2) are the densities of the cluster and the daughter nucleus in a decay, and $v_N(|\mathbf{s}|)$ is the nucleon-nucleon (NN) interaction. The matter density distribution of the daughter or cluster nucleus can be calculated as
\begin{equation}
    \rho(r)=\frac{\rho_{0}}{1+\exp \left(\frac{r-R}{a}\right)}\,,
\end{equation}
where $\rho_0$ is obtained by normalizing $\rho(r)$ to the mass number, $\int\rho(\mathbf{r})\,{\rm d}\mathbf{r}=A$, and the constants are given as $R=1.07A^{1/3}$ fm and $a=0.54$ fm. If the emitted cluster is an $\alpha$ particle, its matter density distribution is given using a standard Gaussian form \cite{Satchler1979}, namely,
\begin{equation}
    \rho_{\alpha}(r) = 0.4229\exp\left(-0.7024\,r^2\right)\,.
\end{equation}

A popular choice of the effective NN interaction is based on the M3Y-Reid-type soft core potential,
\begin{equation}
    v_N\left(|\mathbf{s}|\right)=7999 \frac{\exp \left(-4\left|\mathbf{s}\right|\right)}{4\left|\mathbf{s}\right|}-2134 \frac{\exp \left(-2.5\left|\mathbf{s}\right|\right)}{2.5\left|\mathbf{s}\right|}\,,
\end{equation}
where $|\mathbf{s}|=\left|\mathbf{r}+\mathbf{r}_{c}-\mathbf{r}_{d}\right|$ is the distance between a nucleon in the daughter nucleus and a nucleon in the cluster (Fig. \ref{fig:spherical_nuclei}). This consists of a short-ranged repulsion and a long-ranged attraction responsible for the direct component of the interaction. An additional zero-range contribution $J_{00}\delta(\mathbf{s})$ with $J_{00}=-276(1-0.005\,E/A_c)$ called knock-on exchange term takes into account the antisymmetrization of identical nucleons in the cluster and the daughter nucleus. The latter represents a kind of nonlocality in the DF potential and is usually not included in the calculation if one uses nonlocal nuclear potentials, in order to avoid double counting. Hence, the results in the present work will also be presented without the inclusion of the knock-on exchange term.


As mentioned above, the nuclear potential is modified, multiplying by a constant $\lambda$ such that the Bohr-Sommerfeld quantization condition is satisfied:
\begin{equation}\label{eq:Bohr-Sommerfeld}
    \int_{r_{1}}^{r_{2}} k(r)\,{\rm d} r=\left(n+\frac{1}{2}\right) \pi\,,
\end{equation}
where $k(r)=\sqrt{\frac{2 \mu}{\hbar^{2}}|V(r)-Q|}$ and $n$ is the number of nodes of the quasibound wave function of the cluster nucleus relative motion. This is expressed as $n = (G - l ) /2$, where $G$ is a global quantum number and $l$ is the orbital angular momentum quantum number. The calculated optimal value of $G$ for cluster decay is chosen according to the Wildermuth-Tang condition 
\cite{Wildermuth2013, Ni-Ren2010},
\begin{equation}\label{eq:Wildermuth_rule}
    G=2 n+l=\sum_{i=1}^{A_{c}}\left(g_{i}^{(A_{c}+A_{d})}-g_{i}^{(A_{c})}\right) \,,
\end{equation}
where $g_i^{(Ad+Ac)}$ are the oscillator quantum numbers of the nucleons forming the cluster required to ensure that the cluster is  completely outside the shell occupied by the daughter nucleus, and $g_i^{(Ac)}$ are the internal quantum numbers of the nucleons in the emitted cluster. The values of $g_i$ are taken as $g_i=4$ for nucleons in the $50\leq Z$, $N\leq82$ shell, $g_i=5$ for nucleons in the $82 < Z$, $N\leq126$ shell, $g_i=6$ for nucleons in the $N\leq184$, and $g_i=7$ for nucleons outside the $N=184$ neutron shell closure. These values correspond to the $4\hbar\omega$, $5\hbar\omega$, $6\hbar\omega$, and $7\hbar\omega$ harmonic-oscillator shells, respectively.

The condition in Eq. (\ref{eq:Wildermuth_rule}) ensures that the bound state between the light cluster and the heavy daughter nucleus is an allowed state. It gives us the global quantum number, $G$, that relates the number of nodes, $n$, to the shell model (in this case within a harmonic oscillator basis) and ensures that the Pauli exclusion principle is satisfied \cite{Ni-Ren2010}. In other words, it ensures that the $\alpha$ or light cluster is outside the shell occupied by the daughter. This nonlocality is not the same as that included through the exchange term or the nonlocal framework of the present work. We refer the reader to Refs. \cite{Seif2020,AmirGhodsi} for possible ambiguities introduced by the introduction of this condition in the case of certain kind of potentials.

The Coulomb potential $V_C(r)$ is obtained by using a similar DF procedure with the matter density distributions being replaced by their respective charge density distributions, $\rho^C(r)$. The charge density distributions are taken to have a similar form as that of the matter densities but they are normalized to the number of protons, $\int\rho^{C}(\mathbf{r})\,{\rm d}\mathbf{r}=Z$. The fundamental interaction is the standard proton-proton Coulomb interaction $v_C\left(\mathbf{s}\right)=\frac{e^2}{\left|\mathbf{s}\right|}$. For both nuclear and Coulomb potentials six-dimensional integrals are involved. These are solved numerically by working in momentum space. The detailed procedure can be found in Ref. \cite{Satchler1979}.

\begin{figure*}[ht]
    \includegraphics[scale=0.8]{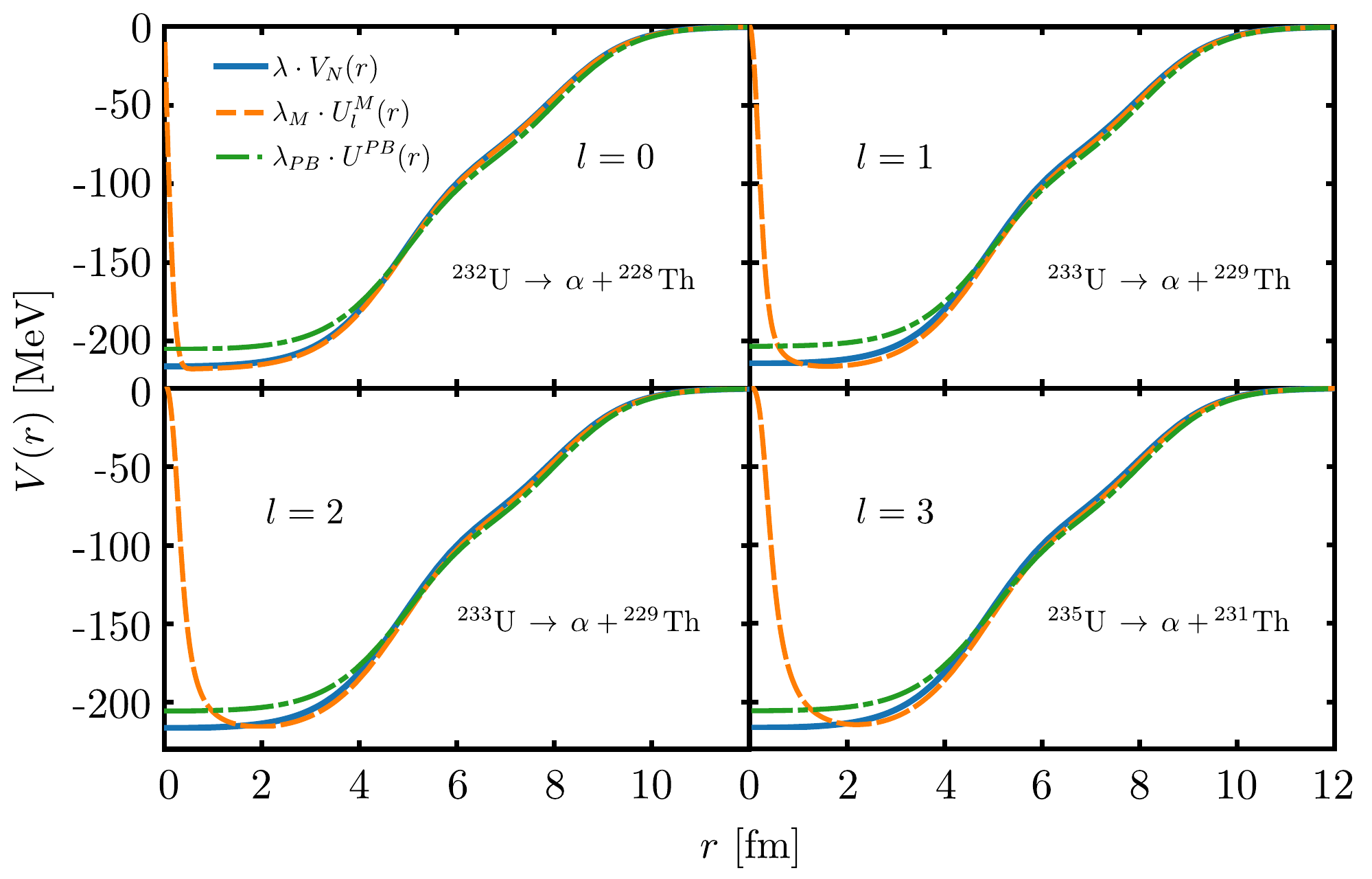}
    \caption{The nuclear interaction potential between $^4$He and different isotopes of thorium (Th) in the $\alpha$ decay of the uranium (U) isotopes. The boxes show a comparison of the effective potentials, $U^{\rm{PB}}(r)$ (dash-dotted line), $U_l^{\rm{M}}(r)$ (dashed line) and the DF potential $V_N(r)$ (solid line) for decays involving different values of the angular momentum quantum number, $l$.}
\label{fig:figure1}
\end{figure*}

\begin{figure*}[ht]
    \includegraphics[scale=0.8]{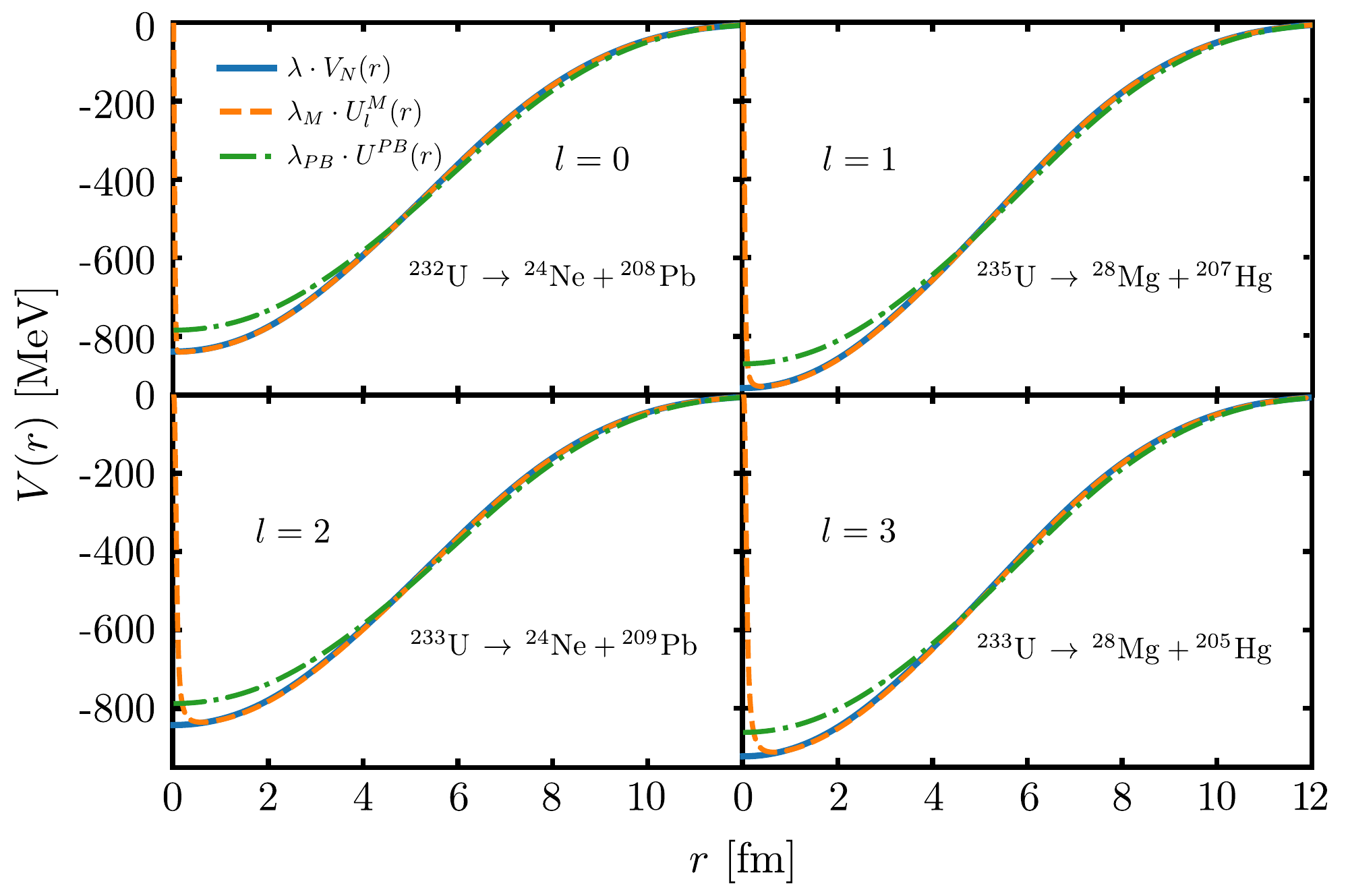}
    \caption{Same as Fig. \ref{fig:figure1} but for the different cluster decay modes of the Uranium (U) isotopes. 
    }\label{fig:figure2}
\end{figure*}

\subsection{Nonlocality}\label{nonlocalformalism}
The existence of nonlocal interactions is a purely quantum mechanical phenomenon. This phenomenon arises because of the very nature of nucleons. When calculating an interaction potential between nuclei, a nonlocality effect appears. This kind of effect has often been discussed in literature in the context of nucleon-nucleus scattering \cite{ahmad, Candido1997, Rotureau}. Taking into account nonlocal approaches improves the description of nuclear interactions, however, introducing nonlocality results in an integro-differential Schr\"odinger equation which can be, in principle, quite difficult to solve. The general form of this equation is written as, 
\begin{multline}\label{eq:nonlocaleq1}
    -{\hbar^2 \over 2 \mu} \nabla^2 \Psi({\bf r}) + [V_{\rm L}({\bf r}) - E]\Psi({\bf r})\\
    = - \int \, {\rm d}{\bf r}^{\prime}\, V_{\rm NL}({\bf r}, {\bf r}^{\prime} ) \Psi({\bf r}^{\prime})\,, 
\end{multline}
where $V_{\rm L}$ can be some isolated local potential and $V_{\rm NL}$ can be the nonlocal one.

The sources of nonlocalities in literature are globally classified into two types: the Feshbach and the Pauli nonlocalities \cite{Balantekin1998}. The Feshbach nonlocality is attributed to inelastic intermediate transitions in scattering processes. In other words, the description of an excitation at a point {\bf r} in space followed by an intermediate state which pro\-pagates and de-excites at some point {\bf r}$^{\prime}$ to get back to the elastic channel (virtual excitations) is contained in the right-hand side of Eq. (\ref{eq:nonlocaleq1}). In this work, we shall not consider the Feshbach nonlocality. The Pauli nonlocality is attributed to the exchange effects which require antisymmetrization of the wave function between the projectile and the target.

In order to solve Eq. (\ref{eq:nonlocaleq1}), it is necessary to know the explicit form of $V_{\rm NL}({\bf r}, {\bf r}^{\prime})$, which as such is not known. However, we do know that $V_{\rm NL}$ should be such, that in the limit of vanishing nonlocality (represented by a nonlocality parameter $b$, with $b \to 0$), the integrodifferential Schr\"odinger equation reduces to the homogeneous one. Assuming this, a nonlocal kernel is usually des\-cribed in literature \cite{PereyBuck1962,Chamon2003,Upadhyay2017} in terms of a factorized form of the potential. The latter is then used to obtain an effective (local) potential which can be used in the homogeneous Schr\"odinger equation.

In this article, we use two nonlocality models to cons\-truct the effective local potential and study the effects of the nonlocality in the $\alpha$ and cluster decay of some heavy nuclei which can decay by both modes. The first approach we consider in which nonlocality is included was proposed by Perey and Buck in the early 1960s \cite{PereyBuck1962}. In the Perey-Buck (PB) model, an energy-independent nonlocal potential for elastic neutron-nucleus scattering is suggested in order to study how far the energy dependence of phenomenological local potentials which had been used earlier could be accounted for by nonlocality. To facilitate calculations, the authors assumed that the nonlocal kernel can be written using a separable form with a local potential times a Gaussian part which contains the nonlocality
\begin{equation}\label{eq:pereybuck1}
    V_{N L}\left(\mathbf{r},\mathbf{r}^{\prime}\right) = \frac{1}{\pi^{3 / 2} b^{3}} U_{N}\left(\frac{1}{2}\left|\mathbf{r}+\mathbf{r}^{\prime}\right|\right)\exp\left[-\left(\frac{\mathbf{r}-\mathbf{r}^{\prime}}{b}\right)^{2}\right]\, , 
\end{equation}
where $b$ is the nonlocality range parameter and $U_N$ is the energy independent potential. The nonlocality range parameter in the case of nucleus-nucleus scattering can be expressed as $b=b_0m_0/\mu$, where $b_0$ is the nonlocal range of the nucleon-nucleus interaction, $m_0$ is the nucleon mass, and $\mu$ is the $\alpha$-daughter or cluster-daughter reduced mass \cite{Jackson1974}. For our calculations, we choose $b_0=0.85$ fm. By using  the factorized form of Eq. (\ref{eq:pereybuck1}), the nonlocal Schr\"odinger equation is given as
\begin{multline}\label{eq:pereybuck2}
    \biggl [{\hbar^2\over 2\mu} \nabla^2 + E \biggr ] \, \Psi_N({\bf r})=\\
    \frac{1}{\pi^{3 / 2} b^{3}}\int U_{N}\left(\frac{1}{2}\left|\mathbf{r}+\mathbf{r}^{\prime}\right|\right)\exp\left[-\left(\frac{\mathbf{r}-\mathbf{r}^{\prime}}{b}\right)^{2}\right] \, {\rm d}{\bf r}^{\prime}\,.
\end{multline}

By assuming that the nonlocal and local wavefunctions are approximately the same [$\Psi_N({\bf r}) \approx \Psi_L({\bf r})$], and the potential inside the nucleus is constant, the authors finally obtained a local equivalent potential given by
\begin{equation}\label{eq:pereybuck8}
    U^{P B}(r) \exp \left\{\frac{\mu b^{2}}{2 \hbar^{2}}\left[E-U^{P B}(r)\right]\right\}=U_{N}(r)\,.
\end{equation}
The $r$ dependence in the equation above is introduced by hand and is justified {\em a posteriori} due to the success of the model in reproducing data. Finally, the transcendental equation [Eq. (\ref{eq:pereybuck8})] is solved to obtain $U^{\rm{PB}}(r)$. The approximations used by the authors introduce an inconsistency at small $r$ \cite{Johan2019}, which as we will see later leads to differences in the results between the two nonlocality models we use.

The second approach in which nonlocality is included was developed in Ref. \cite{Upadhyay2017}. We will refer to this approach as the Mumbai (M) model. In this approach, the authors use the mean value theorem of integral calculus to solve the integrodifferential equation (\ref{eq:nonlocaleq1}). Performing a partial wave expansion of $V_{\rm NL}({\bf r}, {\bf r}^{\prime})$ and $\Psi({\bf r}^{\prime})$ in Eq. (\ref{eq:nonlocaleq1}), they obtained the radial equation, which, in the absence of the spin-orbit and Coulomb interaction was given as 
\begin{multline}\label{eq:nonlocalradial}
    \frac{h^2}{2\mu}\left(\frac{d^2}{dr^2}-\frac{l(l+1)}{r^2}\right)u_l(r)+Eu_l(r)\\
    =\int_0^\infty{g_l(r,r^{\prime})u_l(r^{\prime})\,{\rm d}r^{\prime}}\,,
\end{multline}
where
\begin{multline}\label{eq:gkernel}
    g_{l}\left(r, r^{\prime}\right)=\left(\frac{2 r r^{\prime}}{\sqrt{\pi} b^{3}}\right) \exp \left(-\frac{r^{2}+r^{\prime 2}}{b^{2}}\right)\\
    \times\int_{-1}^{1} U_{N}\left(\frac{\left|\mathbf{r}+\mathbf{r}^{\prime}\right|}{2}\right) \exp \left(\frac{2 r r^{\prime} x}{b^{2}}\right) P_{l}(x)\,{\rm d}x\,.
\end{multline}

By making use of the mean value theorem to rewrite the integral on the right-hand side of Eq. (\ref{eq:nonlocalradial}), and after some simplifications, the authors obtained the homogenized form of the Schr\"odinger equation, 
\begin{equation}
    \left[ \frac{d^2}{dr^2} - \frac{l(l+1)}{r^2} + \frac{2\mu}{\hbar^2} \Big(E  - U_{l}^{M}(r)\Big) \right] u_l(r) = 0\,,
\end{equation}
with an effective strong potential given by \cite{Upadhyay2017}
\begin{equation}\label{eq:mumbaipot}
    U_l^{\rm{M}}(r) = \int_0^{r_m}\, g_l(r,r^{\prime})\,{\rm d}r^{\prime}\, ,
\end{equation}
where $g_l(r,r^{\prime})$ is written as in Eq. (\ref{eq:gkernel}). The upper limit of the integration over $g_l(r,r^{\prime})$ is numerically chosen to be the distance at which the integral becomes negligible, which in our case is essentially the range of the nuclear interaction. In contrast to the PB model, the effective potential obtained in the Mumbai approach is energy independent. However, it is dependent on the angular momentum quantum number, $l$. As a result of these two differences, it shows a different behavior at small distances with $U_l^{\rm{M}} \to 0$ for $r \to 0$. We refer the reader to Ref. \cite{Johan2019} for a detailed discussion of the above differences in the PB and M models. Here we show a comparison of the DF potential $V_N(r)$ and the effective potentials in Figs. (\ref{fig:figure1}) and (\ref{fig:figure2}). To find the PB local equivalent potential $U^{\rm{PB}}(r)$, we replace $U_N(r)$ with $V_N(r)$ and use Eq. (\ref{eq:pereybuck8}). The local effective Mumbai potential [$U_l^{\rm{M}}(r)$] is found using Eq. (\ref{eq:mumbaipot}) by doing a similar replacement: namely, $U_N$ in Eq. (\ref{eq:gkernel}) is replaced by $V_N(r)$. We choose different isotopes of the same nucleus, uranium (U) for comparison in order to see the effect of the $l$ dependence in the potentials. As expected, the behavior of the M and PB potentials is quite different at small distances. Due to the approximation used in the PB model, the potential appears to be almost the same for the different decays and is insensitive to the change in the $l$ value. This pattern is more evident in $\alpha$ decay than in cluster decay. 

A few comments regarding the inclusion of antisymmetrization and medium effects within a nonlocal framework are in order here. The authors in Refs. \cite{Adel2017,adelalpha} have studied the cluster and $\alpha$ decays of several nuclei within a formalism which includes the antisymmetrization through the finite-range exchange part of the nucleon-nucleon interaction \cite{khoaPRC63}. The folding potential in this case becomes energy dependent and becomes nonlocal through the exchange term. The equivalent local potential is obtained by using a localized approximation for the nonlocal densities of the interacting nuclei. One can also include the density dependence of the NN interaction within such a model \cite{khoaPRC56}. As mentioned earlier, in the present work, including antisymmetrization which leads to the so-called ``knock-on exchange" term in the M3Y NN interaction will result in double counting. The effective nonlocal potentials used in this work include the exchange term and medium effects of the NN interaction in a phenomenological way through the nonlocality parameter. In Ref. \cite{chamon}, the authors provided a good description of the nucleus-nucleus scattering data with the S\~ao Paulo potential (which is similar in spirit to the PB potential) {\it without} the exchange term or density dependence included in the NN interaction. The authors noted explicitly that including these effects amounts to double counting of the nonlocality effects and does not provide a good description of data. The nonlocality parameter defined earlier, namely, $b=b_0m_0/\mu$, where $b_0$ is the nonlocal range of the nucleon-nucleus interaction, $m_0$ is the nucleon mass, and $\mu$ is the $\alpha$-daughter or cluster-daughter reduced mass, takes the antisymmetrization and medium effects into account in a phenomenological way.

Finally, we discuss one of the first attempts to eva\-luate the effects of nonlocality in $\alpha$ decay by Chaudhury \cite{chaudhuryPRC,chaudhuryPRL}. The author considered an interaction potential consisting of the pointlike Coulomb interaction between the $\alpha$ and the daughter nucleus superimposed by a nonlocal nuclear potential. The latter was taken to be the Igo potential with the nonlocal part given by a Gaussian function as is used in the approaches mentioned above too. The author obtained a radial differential equation starting from the integro-differential equation (\ref{eq:nonlocaleq1}) with nonlocality and used it to write the JWKB penetration factor. The equation, however, was not the standard radial Schr\"odinger equation, but rather an equation of the form,   
\begin{equation}\label{chaudhury}
    \left[ \frac{d^2}{dr^2} -\frac{l(l+1)}{r^2} + \frac{2 \mu E \,\epsilon(r)}{\hbar^2} - \frac{2 \mu V(r)\, \epsilon(r)}{\hbar^2} \right] \,u_l(r) \, = 0 \,,
\end{equation}
where the factor $\epsilon(r)$ contained the nonlocality parameter $b$. The standard JWKB penetration factor is derived by starting from the radial Schr\"odinger equation for a particle with energy $E$, tunneling a potential barrier $V(r)$. However, the author mentions using a JWKB solution of the above equation to obtain the following preformation factor:
\begin{multline}\label{eq:ChauP1}
    P = \exp\biggl \{ -2 \sqrt{2\mu \over \hbar^2} \int_{r_i}^{r_0} \biggl [ {2(Z-2) e^2 \over r}\,\epsilon(r)\\-\,V_0 f(r)\,\epsilon(r)\,+\,{\hbar^2 \over 2\mu}{l(l+1)\over r^2}\,-\,E\,\epsilon(r) \biggr ]^{1/2} {\rm d}r \biggr \}
\end{multline}
where, $\epsilon(r) = [1 + \eta f(r)]^{-1}$ and $\eta = (\mu b^2 / 2\hbar^2)V_0$. The strong potential, $V_0 f(r)$ was taken from the Igo potential. The penetrability was found to increase by about 50\% due to nonlocality for the nuclei studied. 

\subsection{Potentials for deformed nuclei}\label{deformedpot}

\begin{figure}[h!]
    \includegraphics[scale=1]{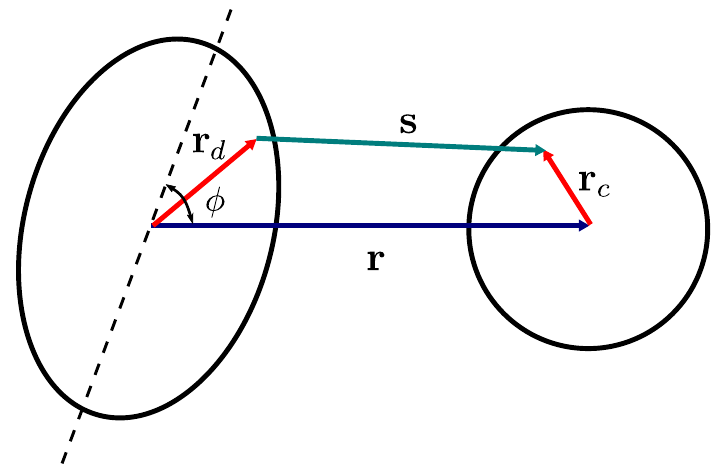}
    \caption{Schematic representation of the spherical-deformed interacting nuclei. $\phi$ is the orientation angle of the symmetry axis of the deformed nucleus with respect to the vector connecting the center of masses of the two nuclei.}
    \label{fig:deformed_nucleus}
\end{figure}
The cluster decay half-lives are calculated using the expression of the decay width given by Eq. (\ref{eq:Gamma}) for the case of spherical nuclei. To extend the cluster decay half-life calculations to deformed nuclei, axial symmetry of nuclei is considered \cite{RhoadesBrown}. An additional simplification of the problem is done by considering just one of the nuclei to be deformed \cite{Maroufi2019_2}. The latter simplification seems to be justified for the decays studied in the present work. Since it is reasonable to consider the $\alpha$ ($^4$He nucleus) to be spherical, in the case of $\alpha$ decay the deformation of the heavy daughter nucleus is taken into account. The hea\-vier daughter in case of cluster decay is often (the doubly magic nucleus) $^{208}$Pb or a nucleus close to it. Thus, in case of cluster decay, we take the deformation of the light cluster into account. This implies the tunneling particle is spherical for $\alpha$ decay and deformed for cluster decay.

For the spherical-deformed interacting system, we begin by writing the nuclear potential between the heavy daughter and the light cluster (or $\alpha$ particle) in Eq. (\ref{eq:V_Double-Folding}) with one spherical density, $\rho_1(r)$, and the other density deformed with axial symmetry, $\rho_2(r,\theta)$. The latter is given as, 


\begin{equation}\label{eq:deformed_density}
    \rho_2\left(r, \theta\right)=\frac{\rho_{0}}{1+\exp \left[\frac{r-R(\theta)}{a(\theta)}\right]}\,,
\end{equation}
where the value of $\rho_0$ is fixed by normalizing the density distribution $\rho(r,\theta)$ and the angle-dependent half-density radius $R(\theta)$ and diffuseness $a(\theta)$ are given by \cite{Bohr-Mottelson1969,Scamps2013, Maroufi2019_2}
\begin{equation}\label{eq:Rtheta}
    R(\theta) = R_0\big(1 +\beta_2Y_{20}(\theta) +\beta_4Y_{40}(\theta)\big)
\end{equation}
and 
\begin{equation}\label{eq:a_deform}
    a(\theta)=a_{\perp}(\theta) \sqrt{1+\left|\vec{\nabla} R(\theta)_{\mid r=R(\theta)}\right|^{2}}\,,
\end{equation}
respectively. In the above expressions $R_0 = 1.07A_d^{1/3}$ fm, $\beta_2$ and $\beta_4$ are quadrupole and hexadecapole deformation parameters of the deformed nucleus, and
\begin{equation}\label{eq:a_perp}
    a_{\perp}(\theta)=a_{0}\big(1+\beta_{2} Y_{20}(\theta)+\beta_{4} Y_{40}(\theta)\big)\,,
\end{equation}
where $a_0$ is taken to be $0.54$ fm. As mentioned earlier, the potential in coordinate space is evaluated by converting the six-dimensional integral in Eq. (\ref{eq:V_Double-Folding}) to a momentum space integral. For the decay problem under consideration, this reduces to a one-dimensional integral in momentum space (see Ref. \cite{Satchler1979} for a detailed derivation with spherical nuclei). When the interacting nuclei are deformed, the shape as well as the orientation of the two nuclei must be taken into account. Assuming one nucleus to be deformed but axially symmetric and the other spherical (for the general case of two deformed nuclei, see Ref. \cite{RhoadesBrown}) the DF nuclear potential can be evaluated as a sum of different multipole components \cite{RhoadesBrown,Ismail2002,Xu-Ren2006} 
\begin{equation}\label{eq:VN_deform}
    V_N(r, \phi)=\sum_{L=0,2,4 \dots} V_N^{L}(r, \phi)\,.
\end{equation}
where $\phi$ is the orientation angle of the symmetry axis of the deformed nucleus with respect to the vector connec\-ting the center of masses of the two nuclei (Fig. \ref{fig:deformed_nucleus}). Each of the multipole component is given by the momentum space integral, 
\begin{multline}\label{eq:deformedpot}
    V_N^{L}(r, \phi) = \frac{2}{\pi}\left[\frac{2L+1}{4\pi}\right]^{1/2}\,P_L(\cos\phi)\;\times\\
    \int_0^{\infty} {\rm d} k\,k^2\,j_L(kr)\rho_1(k)\,\rho_2^L(k)\, v_N(k)\,,
\end{multline}
such that $\rho_2^L(k)$ is the Fourier transform of  
\begin{equation}\label{rhodeformed}
    \rho_2^L(r) = 2 \pi \int \rho_2(r,\theta) \,Y_{L0}(\theta)\,{\rm d}(\cos\theta)\,.
\end{equation}
The above is obtained simply from the multipole expansion 
\begin{equation}\label{eq:rho_expand}
    \rho_2(r, \theta)=\sum_{L=0,2,4} \rho_{L}(r) Y_{L 0}(\theta)\,.
\end{equation}

The sum in Eqs. (\ref{eq:rho_expand}) and (\ref{eq:VN_deform}) is usually truncated at $L=4$ to take the dominant form factors into account. The Bohr–Sommerfeld quantization condition, Eq. (\ref{eq:Bohr-Sommerfeld}), which is an important constraint for the $\alpha$ and cluster decay half-lives calculations (within the JWKB approximation) and determines the depth of the nuclear potential, is now calculated at each angle. This condition is rewritten as
\begin{equation}
    \int_{r_{1}(\phi)}^{r_{2}(\phi)} k(r, \phi)\,{\rm d} r=(G-l+1) \frac{\pi}{2}\,,
\end{equation}
where the turning points are similarly obtained by $V\big(r_i(\phi), \phi\big) = Q$. $l$ is the orbital angular momentum quantum number. The global quantum number $G$ is taken to be the same as in the spherical case and is determined by using the Wildermuth-Tang condition, Eq. (\ref{eq:Wildermuth_rule}). The deformed Coulomb potential is evaluated in a si\-milar manner by using deformed charge density distributions as in Eq. (\ref{eq:deformed_density}). Thus, the total potential is given by
\begin{equation}
    V(r, \phi) = \lambda\, V_{N}(r, \phi)+V_{C}(r,\phi) +\frac{\hbar^{2}}{2\mu}\frac{\left(l+\frac{1}{2}\right)^{2}}{r^{2}}\,.
\end{equation}

As a consequence of this orientation angle dependence of the interaction potential, the wave number is also angle dependent, $k(r,\phi)$. Since this term is required to obtain the decay width $\Gamma$, it is necessary to average over all possible orientation angles of the deformed nucleus. Some authors average the penetration probability and the normalization factor (or the so-called assault frequency of the tunneling particle related to this) and then multiply them \cite{Xu-Ren2006,Adel2017, Maroufi2019_2}. However, it is not appropriate since the penetration probability and the normalization factor are not independent of each other. The suitable way to obtain the decay width is then calculating an angle-dependent width,
\begin{multline}\label{eq:Gamma_phi}
    \Gamma(\phi)=\frac{\hbar^{2}}{2 \mu}\left[\int_{r_{1}(\phi)}^{r_{2}(\phi)} \frac{{\rm d} r}{k(r,\phi)}\right]^{-1}\times\\
    \exp\left[-2 \int_{r_{2}(\phi)}^{r_{3}(\phi)} k(r,\phi)\,{\rm d} r\right]\,,
\end{multline}
and then averaging it over all directions \cite{Denisov2005,Ni-Ren2015,Qian-Ren-Ni2016},
\begin{equation}\label{eq:Gamma_def}
    \Gamma = \frac{1}{2}\int_0^{\pi}\Gamma(\phi)\sin\phi\;{\rm d}\phi\,.
\end{equation}
The $\alpha$ and cluster decay half-lives related to the decay width are calculated using Eqs. (\ref{eq:half-life}), (\ref{eq:Gamma_phi}), and (\ref{eq:Gamma_def}). The preformation factor is chosen to be 1 and will be determined phenomenologically later.

\begin{table*}[h!]
    \centering
    \begin{tabular}{|p{0.08\linewidth}|p{0.12\linewidth}|p{0.09\linewidth}|p{0.08\linewidth}|c|c|p{0.07\linewidth}|p{0.07\linewidth}|p{0.07\linewidth}|p{0.07\linewidth}|}
        \hline\hline
        Parent & Decay mode & \centering $Q$ [MeV] & \centering $b$ [fm] &  $l_{\rm min}$ & \hspace{0.1cm} G \hspace{0.1cm} & \centering $\beta_2^{(c)}$ & \centering $\beta_4^{(c)}$ & \centering $\beta_2^{(d)}$ & \multicolumn{1}{c|}{$\beta_4^{(d)}$} \\\hline\hline
        $^{222}$Ra & $\alpha$+$^{218}$Rn & 6.678 & 0.2160 & 0 & 22 & 0 & 0 & 0.040 & 0.029 \\
        & $^{14}$C+$^{209}$Pb & 33.049 & 0.0648 & 0 & 68 & -0.016 & 0.000 & 0 & 0  \\
        
        &&&&&&&&& \\
        $^{223}$Ra & $\alpha$+$^{219}$Rn & 5.979 & 0.2160 & 2 & 22 & 0 & 0 & 0.103 & 0.072 \\
        & $^{14}$C+$^{209}$Pb & 31.828 & 0.0648 & 4 & 68 & -0.016 & 0.000 & 0 & 0  \\
        
        &&&&&&&&& \\
        $^{228}$Th & $\alpha$+$^{224}$Ra & 5.520 & 0.2160 & 0 & 22 & 0 & 0 & 0.164 & 0.112 \\
        & $^{20}$O+$^{208}$Pb & 44.723 & 0.0466 & 0 & 92 & 0.003 & -0.108 & 0 & 0 \\
        &&&&&&&&& \\ 
        $^{231}$Pa & $\alpha$+$^{227}$Ac & 5.150 & 0.2160 & 0 & 22 & 0 & 0 & 0.172 & 0.112 \\
        & $^{23}$F+$^{208}$Pb & 51.888 & 0.0411 & 1 & 103 & -0.202 & 0.110 & 0 & 0  \\
        
        &&&&&&&&& \\
        $^{232}$U & $\alpha$+$^{228}$Th & 5.413 & 0.2160 & 0 & 22 & 0 & 0 & 0.182 & 0.112  \\
        & $^{24}$Ne+$^{208}$Pb & 62.310 & 0.0395 & 0 & 106 & -0.215 & 0.155 & 0 & 0 \\
        
        &&&&&&&&& \\ 
        $^{233}$U & $\alpha$+$^{229}$Th & 4.908 & 0.2160 & 0 & 22 & 0 & 0 & 0.190 & 0.114 \\
        & $^{24}$Ne+$^{209}$Pb & 60.485 & 0.0395 & 2 & 106 & -0.215 & 0.155 & 0 & 0 \\
        & $^{28}$Mg+$^{205}$Hg & 74.226 & 0.0345 & 3 & 117 & 0.323 & -0.136 & 0 & 0 \\
        
        &&&&&&&&& \\ 
        $^{235}$U & $\alpha$+$^{231}$Th & 4.678 & 0.2160 & 1 & 22 & 0 & 0 &  0.198 & 0.115 \\
        & $^{25}$Ne+$^{210}$Pb & 57.683 & 0.0381 & 3 & 110 & 0 & 0 & 0 & 0 \\
        & $^{28}$Mg+$^{207}$Hg & 72.425 & 0.0345 & 1 & 118 & 0.323 & -0.136 & 0 & 0 \\
        
        &&&&&&&&& \\ 
        $^{237}$Np & $\alpha$+$^{233}$Pa & 4.958 & 0.2160 & 1 & 22 & 0 & 0 & 0.207 & 0.117 \\
        & $^{30}$Mg+$^{207}$Tl & 74.790 & 0.0325 & 2 & 127 & -0.222 & -0.112 & 0 & 0 \\
        
        &&&&&&&&& \\ 
        $^{236}$Pu & $\alpha$+$^{232}$U & 5.867 & 0.2160 & 0 & 22 & 0 & 0 & 0.207 & 0.117 \\
        & $^{28}$Mg+$^{208}$Pb & 79.669 & 0.0345 & 0 & 120 & 0.323 & -0.136 & 0 & 0 \\
        
        &&&&&&&&& \\ 
        $^{241}$Am & $\alpha$+$^{237}$Np & 5.637 & 0.2160 & 1 & 22 & 0 & 0 & 0.215 & 0.102 \\
        & $^{34}$Si+$^{207}$Tl & 93.926 & 0.0291 & 3 & 141 & 0 & 0 & 0 & 0 \\
        
        &&&&&&&&& \\ 
        $^{242}$Cm & $\alpha$+$^{238}$Pu & 6.215 & 0.2160 & 0 & 22 & 0 & 0 & 0.215 & 0.102 \\
        & $^{34}$Si+$^{208}$Pb & 96.509 & 0.0291 & 0 & 142 & 0 & 0 & 0 & 0 \\
        \hline\hline
    \end{tabular}
    \caption{$\alpha$ and cluster decay data. Columns 3 to 6 refer, respectively, to  $Q$-value of the decay from the parent ground state (g.s.) to the daughter g.s., the nonlocality parameter, the angular momentum carried by the emitted particle \cite{NNDC, AMDC}, and Global quantum number given by Eq. (\ref{eq:Wildermuth_rule}). The last columns refer to quadrupole $\beta_2$ and hexadecapole $\beta_4$ deformation parameters of the emitted cluster (c) and daughter nucleus (d), respectively \cite{Moller1995, Sawhney2011}.}
    \label{tab:decay-info}
\end{table*}

\section{Results and discussions}\label{results}
The nonlocal nuclear interaction has been shown to decrease the half-life of spherical nuclei decaying by the emission of $\alpha$ particles \cite{Johan2019}. Here we study the combined effects of nonlocality and deformation of the nuclei involved in the decay process. In Table \ref{tab:decay-info}, $Q$-value of the decay from the parent ground state (g.s.) to the daughter g.s., the nonlocality parameter, the angular momentum carried by the emitted particle \cite{NNDC,AMDC}, and Global quantum number given by Eq. (\ref{eq:Wildermuth_rule}) are listed in columns 3–6. The last columns refer to quadrupole $\beta_2$ and hexadecapole $\beta_4$ deformation parameters of the emitted cluster (c) and daughter nucleus (d), respectively \cite{Moller1995, Sawhney2011}.

\subsection{Half-lives}
In this section, we shall study the effects of nonlocality and deformation individually as well as their joint manifestation in the $\alpha$ and cluster decay half-lives. The calculations are done for two kinds of nonlocal potentials, namely, the energy-independent, so-called Mumbai potential which is $l$ dependent and the well-known Perey-Buck potential which is energy dependent but does not depend on $l$. We define the percentage change (PC) in the half-life of a nucleus as
\begin{equation}\label{eq:PC}
    PC=\frac{t_{1/2}^{\rm DF}-t_{1/2}}{t_{1/2}^{\rm DF}}\times 100,
\end{equation}
where, $t_{1/2}^{\rm DF}$ is the half-life calculated using the DF model only and $t_{1/2}$ is the half-life evaluated within the DF model including nonlocality or deformation or both.

\begin{table*}[h!]
    \centering
    \begin{tabular}{|p{0.08\linewidth}|p{0.12\linewidth}|c|p{0.1\linewidth}|p{0.1\linewidth}|p{0.1\linewidth}|p{0.1\linewidth}|p{0.1\linewidth}|p{0.1\linewidth}|}
        \hline\hline
        & & & \multicolumn{2}{c|}{Double folding} & \multicolumn{2}{c|}{Mumbai} & \multicolumn{2}{c|}{Perey-Buck} \\
        \cline{4 - 5} \cline{6 - 7} \cline{8 - 9}
        Parent & Decay mode & $l_{\rm min}$ & \centering $\lambda$ & \multicolumn{1}{c|}{$t_{1/2}$ [s]} & \centering $\lambda_M$ & \multicolumn{1}{c|}{$PC$} & \centering $\lambda_{PB}$ & \multicolumn{1}{c|}{$PC$} \\\hline\hline
        $^{222}$Ra & $\alpha$+$^{218}$Rn & 0 & 2.04 & 3.28 & 2.05 & 3.2 & 2.32 & 35.8 \\
        & $^{14}$C+$^{208}$Pb & 0 & 1.61 & $1.10\times10^{5}$ & 1.61 & 4.1 & 1.86 & 76.9 \\

        &&&&&&&& \\
        $^{223}$Ra & $\alpha$+$^{219}$Rn & 2 & 2.04 & $7.09\times10^{3}$ & 2.10 & 12.3 & 2.37 & 37.0 \\
        & $^{14}$C+$^{209}$Pb & 4 & 1.61 & $3.25\times10^{7}$ & 1.63 & 17.7 & 1.86 & 77.8 \\
        
        &&&&&&&& \\
        $^{228}$Th & $\alpha$+$^{224}$Ra & 0 & 2.04 & $8.60\times 10^6$ & 2.06 & $3.4$ & 2.37 & $37.5$ \\
        & $^{20}$O+$^{208}$Pb & 0 & 1.49 & $6.78\times10^{12}$ & 1.49 & 3.9 & 1.72 & 87.2 \\
        
        &&&&&&&& \\
        $^{231}$Pa & $\alpha$+$^{227}$Ac & 0 & 2.04 & $4.30\times 10^9$ & 2.06 & $3.2$ & 2.36 & $38.2$ \\
        & $^{23}$F+$^{208}$Pb & 1 & 1.43 & $2.00\times10^{14}$ & 1.44 & 6.4 & 1.66 & 90.2 \\
        
        &&&&&&&& \\ 
        $^{232}$U & $\alpha$+$^{228}$Th & 0 & 2.04 & $3.61\times 10^8$ & 2.05 & $3.2$ & 2.37 & $38.2$ \\
        & $^{24}$Ne+$^{208}$Pb & 0 & 1.42 & $1.43\times10^{10}$ & 1.42 & 3.8 & 1.65 & 91.2 \\
        
        &&&&&&&& \\ 
        $^{233}$U & $\alpha$+$^{229}$Th & 0 & 2.04 & $5.82\times 10^{11}$& 2.06 & $3.3$ & 2.37 & $38.9$ \\
        & $^{24}$Ne+$^{209}$Pb & 2 & 1.42 & $7.25\times10^{12}$ & 1.43 & 10.0 & 1.65 & 91.8 \\
        & $^{28}$Mg+$^{205}$Hg & 3 & 1.35 & $5.80\times10^{13}$ & 1.36 & 13.6 & 1.57 & 94.6 \\
        
        &&&&&&&& \\ 
        $^{235}$U & $\alpha$+$^{231}$Th & 1 & 2.02 & $3.06\times 10^{13}$& 2.07 & $12.2$ & 2.34 & $38.9$ \\
        & $^{25}$Ne+$^{210}$Pb & 3 & 1.41 & $4.57\times10^{17}$ & 1.42 & 13.6 & 1.63 & 92.6 \\
        & $^{28}$Mg+$^{207}$Hg & 1 & 1.36 & $1.10\times10^{16}$ & 1.37 & 6.9 & 1.58 & 94.9 \\
        
        &&&&&&&& \\ 
        $^{237}$Np & $\alpha$+$^{233}$Pa & 1 & 2.01 & $1.11\times 10^{12}$ & 2.06 & $12.4$ & 2.33 & $39.2$ \\
        & $^{30}$Mg+$^{207}$Tl & 2 & 1.35 & $3.11\times10^{14}$ & 1.35 & 9.6 & 1.56 & 94.9 \\
        
        &&&&&&&& \\ 
        $^{236}$Pu & $\alpha$+$^{232}$U & 0 & 2.02 & $9.94\times 10^6$ & 2.04 & $3.3 $ & 2.35 & $38.5$ \\
        & $^{28}$Mg+$^{208}$Pb & 0 & 1.38 & $1.07\times10^{9}$ & 1.38 & 3.7 & 1.60 & 94.0 \\
        
        &&&&&&&& \\ 
        $^{241}$Am & $\alpha$+$^{237}$Np & 1 & 1.99 & $6.55\times 10^8$ & 2.04 & $12.4$ & 2.33 & $38.9$ \\
        & $^{34}$Si+$^{207}$Tl & 3 & 1.32 & $4.33\times10^{10}$ & 1.33 & 12.7 & 1.54 & 96.2 \\
        
        &&&&&&&& \\ 
        $^{242}$Cm & $\alpha$+$^{238}$Pu & 0 & 1.99 & $1.28\times 10^6$ & 2.01 & $3.4 $ & 2.33 & $39.4$ \\
        & $^{34}$Si+$^{208}$Pb & 0 & 1.33 & $9.54\times10^{8}$ & 1.33 & 3.5 & 1.55 & 96.1 \\
        \hline\hline
    \end{tabular}
    \caption{The calculated $\alpha$- and cluster-decay half-lives using the DF model (fifth column) and the percentage decrease in the half-lives of different models of nonlocality compared to those evaluated using the DF model without 
nonlocality. $\lambda$, $\lambda_M$ and $\lambda_{PB}$ are the strengths of 
the nuclear potentials in the DF, M and PB models.}
    \label{tab:nonlocality}
\end{table*}

\subsubsection{Effects of nonlocality}\label{nonlocalresults}
In order to study the change in the half-lives of nuclei due to the nonlocal interaction potential between the decay products, as mentioned before, we use two nonlocal approaches \cite{PereyBuck1962,Upadhyay2017} from literature which are formally very different. We shall see that though both the approaches are equally successful in reproducing the scattering cross-section data, the manifestation of the nonlocal effects in nuclear decay is quite different within the two approaches. In Table \ref{tab:nonlocality}, the half-lives calculated within the DF model are shown for the $\alpha$ and cluster decay modes of different nuclei. The percentage decrease in the half-life due to the inclusion of nonlocality within the Mumbai and Perey-Buck approach is listed in the next columns. A close look at the numbers reveals the differences in the two models which are rooted in the nature of the nonlocality model. The main features to be noted from this table are the following: 
\renewcommand{\labelenumi}{(\roman{enumi})}
\begin{enumerate}
    \item The percentage decrease [listed as PC and evaluated using Eq. (\ref{eq:PC})] is much larger in the Perey-Buck approach as compared to the Mumbai model. 
    \item Whereas the PC remains almost constant (around 38\% decrease for $\alpha$ decay and around 90\% decrease in many cases of cluster decay) within the Perey-Buck approach, this is not the case when the Mumbai model is used.
    \item Within the Mumbai approach, the percentage decrease in half-life increases significantly with increasing value of the relative angular momentum, $l$, between the cluster (or alpha) and the daughter nucleus.
    \item The decrease in half-lives due to nonlocality is always less in the case of $\alpha$ decay as compared to cluster decay of the same nucleus when the Perey-Buck model is used. Such a pattern does not exist if the nonlocal effects are included within the Mumbai model. 
\end{enumerate}

The above features can be understood by noting that (i) whereas the Perey-Buck local equivalent potential, $U^{\rm{PB}}(r)$ is derived starting from the full nonlocal Schr\"odinger equation, the effective Mumbai potential, $U_l^{\rm{M}}(r)$, is derived from the {\it radial} part of the same equation. (ii) As a direct consequence of (i), the Mumbai potential becomes $l$ dependent [see Eqs. (\ref{eq:gkernel}) and (\ref{eq:mumbaipot})]. (iii) As a consequence of starting with the three dimensional Schr\"odinger equation and approximating the nonlocal wave function by the local one, the Perey-Buck potential becomes energy dependent [see Eq. (\ref{eq:pereybuck8})]. (iv) Finally, it must be noted that the derivation of $U^{\rm{PB}}$ is done under the assumption of a constant potential and the $r$ dependence is introduced {\em a posteriori}. The latter gives rise to the biggest differen\-ce between the two nonlocal potentials: namely, the Mumbai potential vanishes near the origin whereas the Perey-Buck potential attains a large constant value.

In Table \ref{tab:nonlocality}, we also list the values of $\lambda$ which decide the strength of the nuclear interaction and are obtained by imposing the Bohr-Sommerfeld condition, Eq. (\ref{eq:Bohr-Sommerfeld}). They remain roughly the same for the decay processes that have the same emitted cluster and do not display any strong $l$ dependence.  

\begin{table*}[tp]
    \centering
    \begin{tabular}{|p{0.06\linewidth}|p{0.13\linewidth}|c|p{0.12\linewidth}|p{0.15\linewidth}|p{0.15\linewidth}|p{0.15\linewidth}|}
        \hline\hline
        & & & \centering $t_{1/2}$ [s] & \multicolumn{3}{c|}{$PC$} \\
        \cline{5 - 7}
        & & & \centering Spherical & \centering Deformed with & \centering Deformed with & \multicolumn{1}{c|}{Deformed with} \\
        Parent & Decay mode & $l_{\rm min}$ & \centering Double & \centering $R(\theta;\beta_2)$ & \centering $R(\theta;\beta_2,\beta_4)$ & \multicolumn{1}{c|}{$R(\theta;\beta_2,\beta_4)$} \\
        & & & \centering folding & \centering $a_{\perp}(\theta;\beta_2)$ & \centering $a_0$ & \multicolumn{1}{c|}{$a_{\perp}(\theta;\beta_2,\beta_4)$} \\
        \hline\hline
        
        $^{222}$Ra & $\alpha$+$^{218}$Rn & 0 & 3.28 & $5.9 $  & $0.4 $ & $8.8 $ \\
        & $^{14}$C+$^{208}$Pb & 0 & $1.10\times10^{5}$ & 0.4 & 0.1 & 0.4 \\
        
        &&&&&& \\
        $^{223}$Ra & $\alpha$+$^{219}$Rn & 2 & $7.09\times 10^3$ & $8.78 $  & $24.8 $ & $33.4 $ \\
        & $^{14}$C+$^{209}$Pb & 4 & $3.25\times10^{7}$ & 0.4 & 0.1 & 0.4 \\
        
        &&&&&& \\
        $^{228}$Th & $\alpha$+$^{224}$Ra & 0 & $8.62\times 10^6$ & $49.1$  & $53.3$ & $66.2$ \\
        & $^{20}$O+$^{208}$Pb & 0 & $6.78\times10^{12}$ & 0.0 & 4.3 & 17.1 \\
        
        &&&&&& \\
        $^{231}$Pa & $\alpha$+$^{227}$Ac & 0 & $4.30\times 10^9$ & $41.3$ & $57.4$ & $68.2$ \\
        & $^{23}$F+$^{208}$Pb & 1 & $2.00\times10^{14}$ & 65.7 & 52.4 & 78.0 \\
        
        &&&&&& \\
        $^{232}$U & $\alpha$+$^{228}$Th & 0 & $3.63\times 10^8$ & $46.8$ & $53.9$ & $62.1$ \\
        & $^{24}$Ne+$^{208}$Pb & 0 & $1.43\times10^{10}$ & 72.0 & 62.7 & 87.5 \\
        
        &&&&&& \\
        $^{233}$U & $\alpha$+$^{229}$Th & 0 & $5.82\times 10^{11}$ & $62.5$ & $48.5$ & $75.4$ \\
        & $^{24}$Ne+$^{209}$Pb & 2 & $7.25\times10^{12}$ & 72.2 & 62.9 & 87.5 \\
        & $^{28}$Mg+$^{205}$Hg & 3 & $5.80\times10^{13}$ & 99.5 & 92.9 & 98.6 \\
        
        &&&&&& \\
        $^{235}$U & $\alpha$+$^{231}$Th & 1 & $3.06\times 10^{13}$ & $45.1$ & $66.1$ & $77.9$ \\
        & $^{28}$Mg+$^{207}$Hg & 1 & $1.10\times10^{16}$ & 99.5 & 93.1 & 98.6 \\
        
        &&&&&& \\
        $^{237}$Np & $\alpha$+$^{233}$Pa & 1 & $1.11\times 10^{12}$ & $51.5$ & $68.8$ & $81.6$ \\
        & $^{30}$Mg+$^{207}$Tl & 2 & $3.11\times10^{14}$ & 83.7 & 63.0 & 80.0 \\
        
        &&&&&& \\
        $^{236}$Pu & $\alpha$+$^{232}$U & 0 & $9.94\times 10^6$ & $49.3$ & $65.9$ & $76.2$ \\
        & $^{28}$Mg+$^{208}$Pb & 0 & $1.07\times10^{9}$ & 99.5 & 92.7 & 98.6\\
        
        &&&&&& \\
        $^{241}$Am & $\alpha$+$^{237}$Np & 1 & $6.55\times 10^8$ & $54.7$ & $34.5$ & $77.6$ \\
        
        &&&&&& \\
        $^{242}$Cm & $\alpha$+$^{238}$Pu & 0 & $1.28\times 10^6$ & $51.9$ & $62.8$ & $81.2$  \\
        \hline\hline
    \end{tabular}
    \caption{The calculated $\alpha$- and cluster-decay half-lives using the DF model (fourth column) and the percentage decrease in the half-lives when different kinds of deformation are included. The fifth column displays the percentage decrease of the half-lives including only the quadrupole deformation parameter $\beta_2$ in both $R(\theta)$ and $a_{\perp}(\theta)$. The sixth column displays the percentage decrease including both the quadrupole and hexadecapole deformation parameters in $R(\theta)$. The diffuseness parameter is constant. The seventh column includes the quadrupole and hexadecapole deformation parameters in both the radius and surface diffuseness parameters.}
    \label{tab:deformation}
\end{table*}

\begin{table*}[tp]
    \centering
    \begin{tabular}{|p{0.06\linewidth}|p{0.11\linewidth}|c|p{0.11\linewidth}|p{0.1\linewidth}|p{0.1\linewidth}|p{0.1\linewidth}|p{0.1\linewidth}|p{0.11\linewidth}|p{0.1\linewidth}|}
        \hline\hline
        & & & \centering $t_{1/2}$ [s] & \multicolumn{6}{c|}{$PC$} \\
        \cline{5 - 10}
        & & & \centering Spherical & \multicolumn{2}{c|}{Deformed with} & \multicolumn{2}{c|}{Deformed with} & \multicolumn{2}{c|}{Deformed with} \\
        Parent & Decay mode & $l_{\rm min}$ & \centering Double & \multicolumn{2}{c|}{$R(\theta;\beta_2)$ and $a_{\perp}(\theta;\beta_2)$} & \multicolumn{2}{c|}{$R(\theta;\beta_2,\beta_4)$ and $a_0$} & \multicolumn{2}{c|}{$R(\theta;\beta_2,\beta_4)$ and $a_{\perp}(\theta;\beta_2,\beta_4)$} \\
        \cline{5 - 6} \cline{7 - 8} \cline{9 - 10}
        & & & \centering folding & \multicolumn{1}{c|}{Mumbai} & \multicolumn{1}{c|}{Perey-Buck} & \multicolumn{1}{c|}{Mumbai} & \multicolumn{1}{c|}{Perey-Buck} & \multicolumn{1}{c|}{Mumbai} & \multicolumn{1}{c|}{Perey-Buck} \\
        \hline\hline
        $^{222}$Ra & $\alpha$+$^{218}$Rn & 0 & 3.28 & 9.1 & 39.6 & 4.2 & 35.9 & 12.3 & 40.8 \\
        & $^{14}$C+$^{209}$Pb & 0 & $1.10\times10^{5}$ & 4.4 & 77.0 & 4.2 & 76.9 & 4.4 & 77.0 \\
        
        & & & & & & & & & \\
        $^{223}$Ra & $\alpha$+$^{219}$Rn & 2 & $7.09\times10^{3}$ & 28.4 & 42.2 & 33.2 & 52.3 & 40.8 & 57.5 \\
        & $^{14}$C+$^{209}$Pb & 4 & $3.25\times10^{7}$ & 18.0 & 77.9 & 17.8 & 77.9 & 18.0 & 77.9 \\
        
        & & & & & & & & & \\
        $^{228}$Th & $\alpha$+$^{224}$Ra & 0 & $8.62\times 10^6$ & $50.3$ & $68.0$ & $54.2$ & $70.8$ & $67.3$ & $79.8$ \\
        & $^{20}$O+$^{208}$Pb & 0 & $6.78\times10^{12}$ & 3.9 & 87.2 & 8.0 & 87.8 & 20.4 & 89.5 \\
        
        & & & & & & & & & \\
        $^{231}$Pa & $\alpha$+$^{227}$Ac & 0 & $4.30\times 10^9$ & $41.5$ & $64.4$ & $58.8$ & $73.8$ & $69.8$ & $83.4$ \\
        & $^{23}$F+$^{208}$Pb & 1 & $2.00\times10^{14}$ & 68.0 & 96.8 & 55.5 & 95.5 & 79.4 & 98.0 \\
        
        & & & & & & & & & \\
        $^{232}$U & $\alpha$+$^{228}$Th & 0 & $3.63\times 10^8$ & $54.1$ & $65.7$ & $55.5$ & $71.6$ & $63.9$ & $77.0$ \\
        & $^{24}$Ne+$^{208}$Pb & 0 & $1.43\times10^{10}$ & 73.0 & 97.7 & 64.2 & 96.9 & 88.0 & 99.0 \\
        
        & & & & & & & & & \\
        $^{233}$U & $\alpha$+$^{229}$Th & 0 & $5.82\times 10^{11}$ & $63.8$ & $76.9$ & $50.3$ & $67.2$ & $76.8$ & $86.2$ \\
        & $^{24}$Ne+$^{209}$Pb & 2 & $7.25\times10^{12}$ & 75.0 & 97.8 & 66.7 & 97.1 & 88.8 & 99.1 \\
        & $^{28}$Mg+$^{205}$Hg & 3 & $5.80\times10^{13}$ & 99.6 & 100.0 & 93.8 & 99.6 & 98.8 & 99.9 \\
        
        & & & & & & & & & \\
        $^{235}$U & $\alpha$+$^{231}$Th & 1 & $3.06\times 10^{13}$ & $62.7$ & $64.5$ & $70.3$ & $80.5$ & $78.3$ & $86.9$ \\
        & $^{25}$Ne+$^{210}$Pb & 3 & $4.57\times10^{17}$ & 13.6 & 92.6 & 13.6 & 92.6 & 13.6 & 92.6 \\
        & $^{28}$Mg+$^{207}$Hg & 1 & $1.10\times10^{16}$ & 99.5 & 100.0 & 93.6 & 99.7 & 98.7 & 99.9 \\
        
        & & & & & & & & & \\
        $^{237}$Np & $\alpha$+$^{233}$Pa & 1 & $1.11\times 10^{12}$ & $63.8$ & $67.7$ & $72.5$ & $81.8$ & $82.9$ & $89.3$ \\
        & $^{30}$Mg+$^{207}$Tl & 2 & $3.11\times10^{14}$ & 85.3 & 99.2 & 66.5 & 98.1 & 81.9 & 99.0 \\
        
        & & & & & & & & & \\
        $^{236}$Pu & $\alpha$+$^{232}$U & 0 & $9.97\times 10^6$ & $53.3$ & $69.4$ & $66.9$ & $79.8$ & $77.3$ & $87.2$\\
        & $^{28}$Mg+$^{208}$Pb & 0 & $1.07\times10^{9}$ & 99.5 & 100.0 & 93.0 & 99.6 & 98.6 & 99.9 \\
        
        & & & & & & & & & \\
        $^{241}$Am & $\alpha$+$^{237}$Np & 1 & $6.55\times 10^8$ & $55.8$ & $73.5$ & $43.2$ & $61.1$ & $78.5$ & $86.1$ \\
        & $^{34}$Si+$^{207}$Tl & 3 & $4.33\times10^{10}$ & 12.7 & 96.2 & 12.7 & 96.2 & 12.7 & 96.2 \\
        
        & & & & & & & & & \\
        $^{242}$Cm & $\alpha$+$^{238}$Pu & 0 & $1.28\times 10^6$ & $54.9$ & $71.0$ & $65.4$ & $77.8$ & $81.9$ & $87.3$\\
        & $^{34}$Si+$^{208}$Pb & 0 & $9.54\times10^{8}$ & 3.5 & 96.1 & 3.5 & 96.1 & 3.5 & 96.1 \\
        \hline\hline
    \end{tabular}
    \caption{The calculated $\alpha$- and cluster-decay half-lives using the DF model and the percentage decrease in half-lives when different models of nonlocality and different kinds of deformation are included. The last two columns represent the complete calculation with $\beta_2$ and $\beta_4$ included in the radius and surface diffuseness parameters. Other columns show the effect of dropping one or both deformation parameters for each model of nonlocality.}
    \label{tab:nonloc_and_deform}
\end{table*}

\subsubsection{Role of deformation}\label{deformresults}
The decay products in the $\alpha$ and cluster decays considered in the present work are not necessarily sphe\-rical nuclei. In the case of $\alpha$ decay, it is the daughter nucleus which can be deformed whereas in case of the cluster decay, it is mostly the light cluster which is deformed. The heavier daughter in this latter case is mostly sphe\-rical. The nuclear deformation is incorporated through a modification of the interaction potential as described in Sec. \ref{deformedpot}. In the previous table, we noticed that for a given parent nucleus and within a given approach for nonlocality, the manifestation of the nonlocal effects in the half-life of the nucleus depends on the mode of decay ($\alpha$ or cluster). In the same spirit, we could also ask if a deformed interaction potential would affect the half-lives of the same parent nucleus decaying by $\alpha$ or cluster decay, in a different way. In a semiclassical scheme as that of the present work, the decay is described by the tunneling of the lighter decay product through the Coulomb barrier created by the interaction between the decay products. Seeing it from this point of view, in the case of $\alpha$ decay, the tunneling nucleus is spherical (and the heavy daughter deformed), whereas in case of cluster decay, it is the deformed light nucleus which tunnels through the barrier generated due to its interaction with the spherical heavier daughter nucleus.

Table \ref{tab:deformation} displays the percentage change in the half-lives of various nuclei decaying by the two modes, namely, $\alpha$ and cluster decay. The last column lists the results with a complete inclusion of the deformation effects in the radius as well as diffuseness parameter in the potential. The fifth and the sixth columns show the effect of ignoring some of the terms involving the hexadecapole and quadrupole deformation (these columns are listed for completeness since such omissions are sometimes found in literature). In spite of the tunneling picture considered above, deformation manifests itself in a similar way in both $\alpha$ and cluster decay. The half-lives decrease after the inclusion of deformation in the interaction potential. The effect is not sensitive to the sign of the deformation parameters but is greater for larger magnitudes of the parameters.

\subsubsection{Half-lives with a nonlocal and deformed interaction potential}\label{nonlocaldeformresults}
In Table \ref{tab:nonloc_and_deform}, we show the effects of including both nonlocality and deformation in the interaction potential between the daughter nuclei. This table is in principle meant to display the overlap of the effects shown in Tables \ref{tab:nonlocality} and \ref{tab:deformation}. However, the overlap of the two effects cannot be expected to be linear. Nonlocality and deformation both  
lead to a decrease in the $\alpha$ as well as cluster decay half-lives. Since the decrease due to nonlocality in the Perey-Buck model is already large, one does not see a significant difference in the results with nonlocality and deformation together. The Mumbai model, however, displayed a smaller decrease in half-lives due to nonlocality and including deformation enhances the effect a lot. The effect is more enhanced for larger values of the angular momentum quantum number $l$. Once again, this difference between the results using the two different approaches for nonlocality could be due to the fact that whereas the Mumbai potential is $l$ dependent but energy independent, the Perey-Buck potential is energy dependent but not $l$ dependent. It is interesting to note that both potentials can reproduce scattering data equally well. The latter is understandable since the differences in the mo\-dels would not show up in the calculation of total cross sections (where we sum over all partial waves). In this sense, the $\alpha$ or the cluster decay of a nucleus which picks up a particular $l$ value can be a better handle to distinguish between different models of nonlocality and fine tune them in a better way.

\subsection{Preformation factors}
\begin{figure*}[ht!]
    \includegraphics[scale=1]{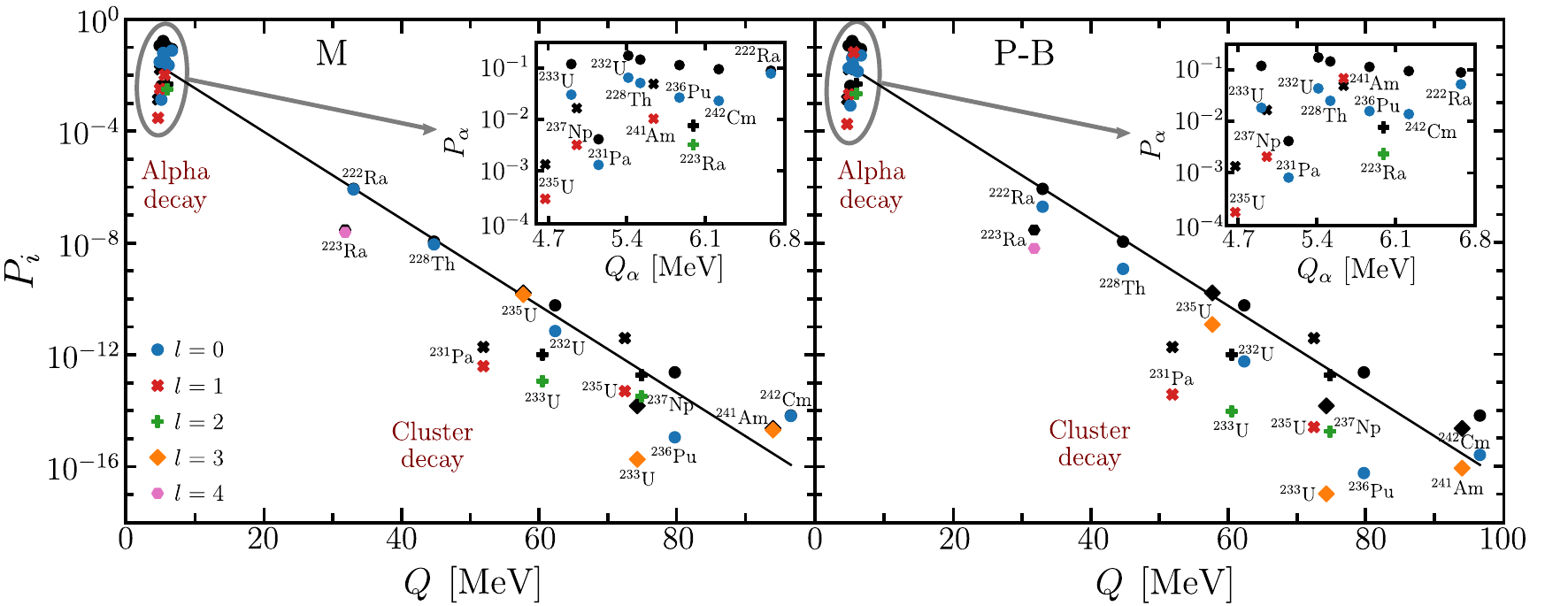}
    \caption{Preformation factors $P_i$ ($i=$ $\alpha$ or $c$) as a function of the $Q$ value or energy of the emitted particle in $\alpha$ and cluster decay of different isotopes. The left panel corresponds to the comparison between the DF model (black symbols) and the Mumbai potential including complete deformation (colored symbols representing different values of $l_{\rm min}$). The right panel compares $P_i$ calculated using the DF and PB models with deformation. The inset shows a magnified view of the $\alpha$ preformation factors which appear to be clubbed together at the small $Q$ values. Most points appear to lie on a band around a straight line which is shown to guide the eye.}\label{fig:figure3}
\end{figure*}
The phenomenon of the emission of $^4$He ($\alpha$) and light nuclei such as $^{14}$C, $^{20}$O, $^{23}$F, $^{24-25}$Ne, $^{28-30}$Mg, and $^{34}$Si in the radioactive decay of nuclei can be explained quite well within a preformed cluster model where the decaying parent nucleus is a cluster of the two decay products \cite{Adel2017,SinghPatraGupta,adelalpha}. The probability of formation of an $\alpha$ particle or a light cluster inside the parent nucleus is proportional to the overlap of the wave function of the final state with that of the initial state. However, one can determine this quantity in a phenomenological way within the preformed cluster model. Within such a model, one usually defines the decay constant or the width as a product of the preformation factor, $P_i$ ($i = \alpha$ or $c$ here), assault frequency, and the penetration probability. Calculations are performed by setting $P_i$ = 1 and the disagreement with the experimental half-life is then attributed to the performation probability. Thus, one evaluates the theoretical width $\Gamma$ as in Eq. (\ref{eq:Gamma}), or Eq. (\ref{eq:Gamma_def}), with $P_i$ = 1, and then defines the probability for cluster formation as \cite{SinghPatraGupta}, $P_i = \Gamma^{\rm exp} / \Gamma$. Here $\Gamma^{\rm exp} = \hbar \ln 2 / t_{1/2}^{\rm exp}$ is given by the experimental half-life. Note that in case the parent nucleus decays into more than one channel, one must take the branching ratio for the channel under consideration into account. For decays involving the parents as well as daughter nuclei with non zero spins, the $\alpha$ or the light cluster can be emitted with a non-zero angular momentum quantum number. In this case, the width calculated using Eq. (\ref{eq:Gamma}) is in principle a partial width, $\Gamma_l$. In order to compare with the experimental (full) width, one would in principle have to consider the sum $\sum_l P_i^{(l)}\, \Gamma_{l} = \Gamma^{\rm exp}$. Indeed, in microscopic calculations, one does consider the so-called ``reduced widths" and evaluates the $l$-dependent preformation factor within a given theoretical model. One of the first such calculations was performed by Mang \cite{mang1, mang2} in a test of the shell model with the Po and At isotopes. For a phenomenological determination of the preformation factor within the preformed cluster model, however, it is customary to define $P_i$ as
\begin{equation}\label{pfactor}
    P_i = \frac{\Gamma^{\rm exp}}{\Gamma_{l_{\rm min}}} 
\end{equation}
with $\Gamma_{l_{\rm min}}$ evaluated for the minimum $l$ value 
using Eq. (\ref{eq:Gamma}) with $P_i$ = 1. The above relation is a good approxi\-mation if the preformation factors and widths for all $l > l_{\rm min}$ are much smaller than those for $l_{\rm min}$. In the absence of any theoretical calculations of $P_i$ confirming the latter, we shall evaluate the preformation factors u\-sing Eq. (\ref{pfactor}), which is commonly found in literature. In the tables \ref{tab:Sc_nonloc_and_deform} and \ref{tab:Sc_nonloc_plus_deform}, we present the effects of nonlocality and deformation on the values of $P_i$ for the different decays discussed earlier. The notation is similar to that used in the tables showing the half-lives and the percentage changes due to nonlocality and deformation. We note that the preformation factors in the tables have been evaluated only for the transitions from the parent ground state (g.s) to the g.s. of the heavy daughter nucleus using Eq. (\ref{pfactor}). The branching ratio (B.R.) for the transition from (parent) g.s. $\to$ (daughter) g.s. is listed in the table in case of $\alpha$ decay and should be multiplied with the values of $P_i$ given in the table in order to determine the performation probability for this particular transition. Thus, the preformation factors for transitions from the parent g.s. to the different excited states of the daughter will differ depending on the B.R. to that level. Since the focus of the present work is to investigate the effects of nonlocality and deformation, we leave such an investigation of g.s. to excited states transitions for the future.

Though the tables give all details of the dependence of the preformation factors on deformation and nonlocality (individually as well as the combined effect), in Fig. \ref{fig:figure3}, we plot the values of $P_i$ for $\alpha$ and cluster decay as a function of the $Q$ values in order to visualize these effects. The two panels show the calculations with two different models of nonlocality with the inclusion of deformation with all the relevant terms in the radius and the diffuseness parameters. The insets show a magnified view of the preformation factors, $P_i$ (with $i$ = $\alpha$) for $\alpha$ decay which appear in the left corner of each panel. Different interesting features emerge from this figure:
\renewcommand{\labelenumi}{(\roman{enumi})}
\begin{enumerate}
    \item All values of $P_i$ evaluated for different isotopes seem to lie on a band, displaying a linear dependence as a function of energy.
    \item  For a given energy, the preformation factors decrease as a function of $l$ (shown by the colored symbols).
    \item  The effect of nonlocality + deformation is more pronounced in case of the Perey-Buck (PB) model (right panel) as compared to the Mumbai model (left panel). The black symbols represent the calculations within the double folding (DF) model without the inclusion of nonlocality or deformation.
\end{enumerate}

The $Q$ values or the energies carried by the $\alpha$ or light cluster are related to the binding energies of the parent and daughter nuclei which in turn also define the formation energy of a cluster \cite{DengRen}. A bigger binding energy of the daughter therefore implies a higher $Q$ value and more internal energy available for the two nuclei in the cluster. The latter explains the decrease in the values of $P_i$ with increasing values of $Q$. The observation of a li\-near dependence of the phenomenologically evaluated $P_i$ in Fig. \ref{fig:figure3} can be useful for theoretical investigations of cluster formation probabilities. 
\begin{table*}[h!]
    \centering
    \begin{tabular}{|p{0.06\linewidth}|p{0.1\linewidth}|p{0.05\linewidth}|c|p{0.11\linewidth}|p{0.11\linewidth}|p{0.11\linewidth}|p{0.11\linewidth}|p{0.11\linewidth}|p{0.11\linewidth}|}
        \hline\hline
        & & & & \multicolumn{6}{c|}{$P_i$} \\
        \cline{5 - 10}
        & & & & \centering Spherical & \multicolumn{2}{c|}{Nonlocality} & \multicolumn{3}{c|}{Deformation} \\
        \cline{6 - 7} \cline{8 - 10}
        Parent & Decay mode & B.R. & $l_{\rm min}$ & \centering Double & \multicolumn{1}{c|}{Mumbai} & \multicolumn{1}{c|}{Perey-Buck} & \centering $R(\theta;\beta_2)$ & \centering $R(\theta;\beta_2,\beta_4)$ & \multicolumn{1}{c|}{$R(\theta;\beta_2,\beta_4)$} \\
        & & & & \centering Folding & & & \centering $a_{\perp}(\theta;\beta_2)$ & \centering $a_0$ & \multicolumn{1}{c|}{$a_{\perp}(\theta;\beta_2,\beta_4)$} \\
        \hline\hline
        $^{222}$Ra & $\alpha$+$^{218}$Rn & 0.969 & 0 & 8.63$\times 10^{-2}$ & 8.35$\times 10^{-2}$ & 5.54$\times 10^{-2}$ & 8.11$\times 10^{-2}$ & 8.59$\times 10^{-2}$ & 7.82$\times 10^{-2}$\\
        & $^{14}$C+$^{208}$Pb & - & 0 & 8.64$\times 10^{-7}$ & 8.29$\times 10^{-7}$ & 2.00$\times 10^{-7}$ & 8.61$\times10^{-7}$ & 8.63$\times10^{-7}$ & 8.61$\times10^{-7}$ \\
        
        &&&&&&&&&\\
        $^{223}$Ra & $\alpha$+$^{219}$Rn & 0.01 & 2 & 7.18$\times 10^{-3}$ & 6.30$\times 10^{-3}$ & 5.42$\times 10^{-3}$ & 6.55$\times 10^{-3}$ & 5.40$\times 10^{-3}$ & 4.82$\times 10^{-3}$ \\
        & $^{14}$C+$^{209}$Pb & - & 4 & 2.93$\times 10^{-8}$ & 2.41$\times 10^{-8}$ & 6.50$\times 10^{-9}$ & 2.92$\times10^{-8}$ & 2.92$\times10^{-8}$ & 2.92$\times10^{-8}$ \\
        
        &&&&&&&&&\\
        $^{228}$Th & $\alpha$+$^{224}$Ra & 0.734 & 0 & 1.40$\times 10^{-1}$ & 1.37$\times 10^{-1}$ & 8.37$\times 10^{-2}$ & 7.28$\times 10^{-2}$ & 6.68$\times 10^{-2}$ & 5.23$\times 10^{-2}$ \\
        & $^{20}$O+$^{208}$Pb & - & 0 & 1.12$\times 10^{-8}$ & 1.08$\times 10^{-8}$ & 1.44$\times 10^{-9}$ & 1.12$\times10^{-8}$ & 1.08$\times10^{-8}$ & 9.31$\times10^{-9}$ \\
        
        &&&&&&&&&\\
        $^{231}$Pa & $\alpha$+$^{227}$Ac & 0.110 & 0 & 4.16$\times 10^{-3}$ & 4.03$\times 10^{-3}$ & 2.57$\times 10^{-3}$ & 2.44$\times 10^{-3}$ & 1.78$\times 10^{-3}$ & 1.38$\times 10^{-3}$ \\
        & $^{23}$F+$^{208}$Pb & - & 1 & 1.91$\times10^{-12}$ & 1.79$\times10^{-12}$ & 1.87$\times10^{-13}$ & 6.56$\times10^{-13}$ & 9.11$\times10^{-13}$ & 4.21$\times10^{-13}$ \\
        
        &&&&&&&&&\\
        $^{232}$U & $\alpha$+$^{228}$Th & 0.685 & 0 & 1.67$\times 10^{-1}$ & 1.61$\times 10^{-1}$ & 1.02$\times 10^{-1}$ & 8.88$\times 10^{-2}$ & 7.69$\times 10^{-2}$ & 6.53$\times 10^{-2}$ \\
        & $^{24}$Ne+$^{208}$Pb & - & 0 & 5.91$\times10^{-11}$ & 1.65$\times10^{-11}$ & 5.20$\times10^{-12}$ & 1.48$\times10^{-10}$ & 2.20$\times10^{-11}$ & 7.39$\times10^{-12}$ \\
        
        &&&&&&&&&\\
        $^{233}$U & $\alpha$+$^{229}$Th & 0.843 & 0 & 1.15$\times 10^{-1}$ & 1.12$\times 10^{-1}$ & 7.08$\times 10^{-2}$ & 4.42$\times 10^{-2}$ & 5.95$\times 10^{-2}$ & 3.08$\times 10^{-2}$ \\
        & $^{24}$Ne+$^{209}$Pb & - & 2 & 1.04$\times10^{-12}$ & 9.36$\times10^{-13}$ & 9.57$\times10^{-14}$ & 2.89$\times10^{-13}$ & 1.99$\times10^{-12}$ & 1.30$\times10^{-13}$ \\
        & $^{28}$Mg+$^{205}$Hg & - & 3 & 1.50$\times10^{-14}$ & 1.30$\times10^{-14}$ & 8.12$\times10^{-16}$ & 7.78$\times10^{-17}$ & 1.07$\times10^{-15}$ & 2.12$\times10^{-16}$ \\
        
        &&&&&&&&&\\
        $^{235}$U & $\alpha$+$^{231}$Th & 0.0477 & 1 & 1.37$\times 10^{-3}$ & 1.21$\times 10^{-3}$ & 8.42$\times 10^{-4}$ & 7.56$\times 10^{-4}$ & 4.67$\times 10^{-4}$ & 3.23$\times 10^{-4}$ \\
        & $^{25}$Ne+$^{210}$Pb & - & 3 & 1.65$\times10^{-10}$ & 1.42$\times10^{-10}$ & 1.22$\times10^{-11}$ & 1.65$\times10^{-10}$ & 1.65$\times10^{-10}$ & 1.65$\times10^{-10}$ \\
        & $^{28}$Mg+$^{207}$Hg & - & 1 & 3.97$\times10^{-12}$ & 3.70$\times10^{-12}$ & 2.03$\times10^{-13}$ & 1.98$\times10^{-14}$ & 2.74$\times10^{-13}$ & 5.41$\times10^{-14}$ \\
        
        &&&&&&&&&\\
        $^{237}$Np & $\alpha$+$^{233}$Pa & 0.0239 & 1 & 1.64$\times 10^{-2}$ & 1.44$\times 10^{-2}$ & 1.00$\times 10^{-2}$ & 8.00$\times 10^{-3}$ & 5.14$\times 10^{-2}$ & 3.71$\times 10^{-2}$ \\
        & $^{30}$Mg+$^{207}$Tl & - & 2 & 1.84$\times10^{-13}$ & 1.66$\times10^{-13}$ & 9.37$\times10^{-15}$ & 2.99$\times10^{-14}$ & 6.80$\times10^{-14}$ & 3.68$\times10^{-14}$ \\
        
        &&&&&&&&&\\
        $^{236}$Pu & $\alpha$+$^{232}$U & 0.691 & 0 & 1.10$\times 10^{-1}$ & 1.06$\times 10^{-1}$ & 6.80$\times 10^{-2}$ & 5.60$\times 10^{-2}$ & 3.77$\times 10^{-2}$ & 2.78$\times 10^{-2}$ \\
        & $^{28}$Mg+$^{208}$Pb & - & 0 & 2.38$\times10^{-13}$ & 2.29$\times10^{-13}$ & 1.43$\times10^{-14}$ & 1.17$\times10^{-15}$ & 1.74$\times10^{-14}$ & 4.02$\times10^{-13}$ \\
        
        &&&&&&&&&\\
        $^{241}$Am & $\alpha$+$^{237}$Np & 0.0037 & 1 & 4.80$\times 10^{-2}$ & 4.20$\times 10^{-2}$ & 2.93$\times 10^{-2}$ & 2.22$\times 10^{-2}$ & 3.09$\times 10^{-2}$ & 1.18$\times 10^{-2}$ \\
        & $^{34}$Si+$^{207}$Tl & - & 3 & 2.35$\times10^{-15}$ & 2.05$\times10^{-15}$ & 8.92$\times10^{-17}$ & 2.35$\times10^{-15}$ & 2.35$\times10^{-15}$ & 2.35$\times10^{-15}$ \\
        
        &&&&&&&&&\\
        $^{242}$Cm & $\alpha$+$^{238}$Pu & 0.748 & 0 & 9.16$\times 10^{-2}$ & 8.85$\times 10^{-2}$ & 5.52$\times 10^{-2}$ & 4.40$\times 10^{-2}$ & 3.40$\times 10^{-2}$ & 1.88$\times 10^{-2}$ \\
        & $^{34}$Si+$^{208}$Pb & - & 0 & 6.78$\times10^{-15}$ & 6.55$\times10^{-15}$ & 2.65$\times10^{-16}$ & 6.78$\times10^{-15}$ & 6.78$\times10^{-15}$ & 6.78$\times10^{-15}$ \\
        \hline\hline
    \end{tabular}
    \caption{Preformation factor $P_i$ ($i$ = $\alpha$ or $c$) using the DF model (fourth column), different models of nonlocality (fifth and sixth columns), and including different kinds of deformation (seventh, eighth, and ninth columns). The last column represents the calculation with $\beta_2$ and $\beta_4$ included in the radius and diffuseness parameters. Since the calculations are performed for transitions from the parent to the daughter ground state (g.s), the branching ratio (B.R.) for the g.s. $\to$ g.s. transition, whenever available from experiment is given.}
    \label{tab:Sc_nonloc_and_deform}
\end{table*}

\begin{table*}[tp]
    \centering
    \begin{tabular}{|p{0.06\linewidth}|p{0.1\linewidth}|p{0.05\linewidth}|c|p{0.11\linewidth}|p{0.1\linewidth}|p{0.1\linewidth}|p{0.1\linewidth}|p{0.1\linewidth}|p{0.11\linewidth}|p{0.1\linewidth}|}
        \hline\hline
        & & & & \multicolumn{7}{c|}{$P_i$} \\
        \cline{5 - 11}
        & & & & \centering Spherical & \multicolumn{2}{c|}{Deformed with} & \multicolumn{2}{c|}{Deformed with} & \multicolumn{2}{c|}{Deformed with} \\
        Parent & Decay mode & B.R. & $l_{\rm min}$ & \multicolumn{1}{c|}{Double} & \multicolumn{2}{c|}{$R(\theta;\beta_2)$ and $a_{\perp}(\theta;\beta_2)$} & \multicolumn{2}{c|}{$R(\theta;\beta_2,\beta_4)$ and $a_0$} & \multicolumn{2}{c|}{$R(\theta;\beta_2,\beta_4)$ and $a_{\perp}(\theta;\beta_2,\beta_4)$} \\
        \cline{6 - 7} \cline{8 - 9} \cline{10 - 11}
        & & & & \multicolumn{1}{c|}{Folding} & \multicolumn{1}{c|}{Mumbai} & \multicolumn{1}{c|}{Perey-Buck} & \multicolumn{1}{c|}{Mumbai} & \multicolumn{1}{c|}{Perey-Buck} & \multicolumn{1}{c|}{Mumbai} & \multicolumn{1}{c|}{Perey-Buck} \\
        \hline\hline
        
        $^{222}$Ra & $\alpha$+$^{218}$Rn & 0.969 & 0 & 8.63$\times10^{-2}$ & 7.84$\times10^{-2}$ & 5.20$\times10^{-2}$ & 8.26$\times10^{-2}$ & 5.52$\times10^{-2}$ & 7.58$\times10^{-2}$ & 5.10$\times 10^{-2}$ \\
        & $^{14}$C+$^{208}$Pb & - & 0 & 8.64$\times10^{-7}$ & 8.26$\times10^{-7}$ & 1.99$\times10^{-7}$ & 8.28$\times10^{-7}$ & 2.00$\times10^{-7}$ & 8.26$\times10^{-7}$ & 1.99$\times10^{-7}$ \\
        
        & & & & & & & & & & \\
        $^{223}$Ra & $\alpha$+$^{219}$Rn & 0.01 & 2 & 7.18$\times10^{-3}$ & 5.14$\times10^{-3}$ & 4.15$\times10^{-3}$ & 4.80$\times10^{-3}$ & 3.42$\times10^{-3}$ & 4.25$\times10^{-3}$ & 3.05$\times10^{-3}$ \\
        & $^{14}$C+$^{209}$Pb & - & 4 & 2.93$\times10^{-8}$ & 2.40$\times10^{-8}$ & 6.46$\times10^{-9}$ & 2.41$\times10^{-8}$ & 6.48$\times10^{-9}$ & 2.40$\times10^{-8}$ & 6.46$\times10^{-9}$ \\
        
        & & & & & & & & & & \\
        $^{228}$Th & $\alpha$+$^{224}$Ra & 0.734 & 0 & 1.40$\times 10^{-1}$ & 7.21$\times 10^{-2}$ & 4.57$\times 10^{-2}$ & 6.54$\times 10^{-2}$ & 4.61$\times 10^{-2}$ & 4.99$\times 10^{-2}$ & 2.46$\times 10^{-2}$ \\
        & $^{20}$O+$^{208}$Pb & - & 0 & 1.12$\times10^{-8}$ & 1.08$\times10^{-8}$ & 1.44$\times10^{-9}$ & 1.03$\times10^{-8}$ & 1.38$\times10^{-9}$ & 8.95$\times10^{-9}$ & 1.19$\times10^{-9}$ \\
        
        & & & & & & & & & & \\
        $^{231}$Pa & $\alpha$+$^{227}$Ac & 0.110 & 0 & 4.16$\times 10^{-3}$ & 2.43$\times 10^{-3}$ & 1.48$\times 10^{-3}$ & 1.71$\times 10^{-3}$ & 1.09$\times 10^{-3}$ & 1.34$\times 10^{-3}$ & 8.30$\times 10^{-4}$ \\
        & $^{23}$F+$^{208}$Pb & - & 1 & 1.91$\times10^{-12}$ & 6.13$\times10^{-13}$ & 6.18$\times10^{-14}$ & 8.51$\times10^{-13}$ & 8.65$\times10^{-14}$ & 3.93$\times10^{-13}$ & 3.84$\times10^{-14}$ \\
        
        & & & & & & & & & & \\
        $^{232}$U & $\alpha$+$^{228}$Th & 0.685 & 0 & 1.67$\times 10^{-1}$ & 7.66$\times 10^{-2}$ & 5.73$\times 10^{-2}$ & 7.61$\times 10^{-2}$ & 4.73$\times 10^{-2}$ & 6.31$\times 10^{-2}$ & 4.25$\times 10^{-2}$ \\
        & $^{24}$Ne+$^{208}$Pb & - & 0 & 5.91$\times10^{-11}$ & 1.59$\times10^{-11}$ & 1.38$\times10^{-12}$ & 2.12$\times10^{-11}$ & 1.84$\times10^{-12}$ & 7.11$\times10^{-12}$ & 5.86$\times10^{-13}$ \\
        
        & & & & & & & & & & \\
        $^{233}$U & $\alpha$+$^{229}$Th & 0.843 & 0 & 1.15$\times 10^{-1}$ & 5.68$\times 10^{-2}$ & 2.67$\times 10^{-2}$ & 5.76$\times 10^{-2}$ & 3.80$\times 10^{-2}$ & 2.98$\times 10^{-2}$ & 1.81$\times 10^{-2}$ \\
        & $^{24}$Ne+$^{209}$Pb & - & 2 & 1.04$\times10^{-12}$ & 2.60$\times10^{-13}$ & 2.27$\times10^{-14}$ & 3.46$\times10^{-13}$ & 3.03$\times10^{-14}$ & 1.17$\times10^{-13}$ & 9.69$\times10^{-15}$ \\
        & $^{28}$Mg+$^{205}$Hg & - & 3 & 1.50$\times10^{-14}$ & 6.71$\times10^{-14}$ & 3.49$\times10^{-15}$ & 9.29$\times10^{-16}$ & 5.64$\times10^{-17}$ & 1.84$\times10^{-16}$ & 1.08$\times10^{-17}$ \\
        
        & & & & & & & & & & \\
        $^{235}$U & $\alpha$+$^{231}$Th & 0.0477 & 1 & 1.37$\times 10^{-3}$ & 5.14$\times 10^{-4}$ & 4.88$\times 10^{-4}$ & 4.09$\times 10^{-4}$ & 2.67$\times 10^{-4}$ & 2.99$\times 10^{-4}$ & 1.80$\times 10^{-4}$ \\
        & $^{25}$Ne+$^{210}$Pb & - & 3 & 1.65$\times10^{-10}$ & 1.42$\times10^{-10}$ & 1.22$\times10^{-11}$ & 1.42$\times10^{-10}$ & 1.22$\times10^{-11}$ & 1.42$\times10^{-10}$ & 1.22$\times10^{-11}$ \\
        & $^{28}$Mg+$^{207}$Hg & - & 1 & 3.97$\times10^{-12}$ & 1.84$\times10^{-14}$ & 8.50$\times10^{-16}$ & 2.56$\times10^{-13}$ & 1.38$\times10^{-14}$ & 5.05$\times10^{-14}$ & 2.63$\times10^{-15}$ \\
        
        & & & & & & & & & & \\
        $^{237}$Np & $\alpha$+$^{233}$Pa & 0.0239 & 1 & 1.64$\times 10^{-2}$ & 5.94$\times 10^{-3}$ & 4.31$\times 10^{-3}$ & 4.52$\times 10^{-3}$ & 3.00$\times 10^{-3}$ & 3.24$\times 10^{-3}$ & 2.09$\times 10^{-3}$ \\
        & $^{30}$Mg+$^{207}$Tl & - & 2 & 1.84$\times10^{-13}$ & 2.70$\times10^{-14}$ & 1.45$\times10^{-15}$ & 6.16$\times10^{-14}$ & 3.46$\times10^{-15}$ & 3.33$\times10^{-14}$ & 1.85$\times10^{-15}$ \\
        
        & & & & & & & & & & \\
        $^{236}$Pu & $\alpha$+$^{232}$U & 0.691 & 0 & 1.10$\times 10^{-1}$ & 6.27$\times 10^{-2}$ & 3.38$\times 10^{-2}$  & 3.65$\times 10^{-2}$ & 2.28$\times 10^{-2}$ & 2.62$\times 10^{-2}$ & 1.56$\times 10^{-2}$  \\
        & $^{28}$Mg+$^{208}$Pb & - & 0 & 2.38$\times10^{-13}$ & 1.13$\times10^{-15}$ & 5.94$\times10^{-17}$ & 1.67$\times10^{-14}$ & 1.03$\times10^{-15}$ & 3.88$\times10^{-13}$ & 2.81$\times10^{-14}$ \\
        
        & & & & & & & & & & \\
        $^{241}$Am & $\alpha$+$^{237}$Np & 0.0037 & 1 & 4.80$\times 10^{-2}$ & 2.24$\times 10^{-2}$ & 1.27$\times 10^{-2}$ & 2.69$\times 10^{-2}$ & 1.86$\times 10^{-2}$ & 1.03$\times 10^{-2}$ & 6.65$\times 10^{-3}$ \\
        & $^{34}$Si+$^{207}$Tl & - & 3 & 2.35$\times10^{-15}$ & 2.05$\times10^{-15}$ & 8.92$\times10^{-17}$ & 2.05$\times10^{-15}$ & 8.92$\times10^{-17}$ & 2.05$\times10^{-15}$ & 8.92$\times10^{-17}$ \\
        
        & & & & & & & & & & \\
        $^{242}$Cm & $\alpha$+$^{238}$Pu & 0.748 & 0 & 9.16$\times 10^{-2}$ & 3.90$\times 10^{-2}$ & 2.64$\times10^{-2}$  & 3.16$\times 10^{-2}$ & 2.03$\times 10^{-2}$ & 2.27$\times 10^{-2}$ & 1.37$\times 10^{-2}$  \\
        & $^{34}$Si+$^{208}$Pb & - & 0 & 6.78$\times10^{-15}$ & 6.55$\times10^{-15}$ & 2.65$\times10^{-16}$ & 6.55$\times10^{-15}$ & 2.65$\times10^{-16}$ & 6.55$\times10^{-15}$ & 2.65$\times10^{-16}$ \\
        \hline\hline
    \end{tabular}
    \caption{The calculated $\alpha$- and cluster-decay preformation factor $P_i$ ($i$ = $\alpha$ or $c$) using the DF model and different models of nonlocality (including deformation partly and fully). The last two columns represent the complete calculation with $\beta_2$ and $\beta_4$ included in the radius and diffuseness parameters. Other columns show the effect of dropping one or both deformation parameters. B.R. is the branching ratio for the transition from the parent ground state to the daughter ground state as in \ref{tab:Sc_nonloc_and_deform}.}
    \label{tab:Sc_nonloc_plus_deform}
\end{table*}

\subsection{Comparison of penetration probabilities}

\begin{table*}[h!]
    \centering
    \begin{tabular}{|p{0.08\linewidth}|p{0.12\linewidth}|p{0.07\linewidth}|c|p{0.11\linewidth}|p{0.07\linewidth}|p{0.07\linewidth}|}
        \hline\hline
        Parent & Decay mode & \centering $Q$ [MeV] & \centering $l_{\rm min}$ & \centering $P^{(DF)}$ & \centering $PC^{(M)}$ & \multicolumn{1}{c|}{$PC^{(Ch)}$} \\
        \hline\hline
        $^{222}$Ra & $\alpha$+$^{218}$Rn & 6.678  & 0 & $7.40\times 10^{-23}$ & 3.9 & 7.1 \\
        & $^{14}$C+$^{208}$Pb & 33.049 & 0 & $3.23\times10^{-27}$ & 4.1 & 7.9 \\
        
        &&&&&& \\
        $^{223}$Ra & $\alpha$+$^{219}$Rn & 5.979  & 2 & $3.41\times 10^{-26}$ & 12.2 & 7.4 \\
        & $^{14}$C+$^{209}$Pb & 31.828 & 4 & $1.08\times10^{-29}$ & 21.1 & 8.3 \\
        
        &&&&&& \\
        $^{228}$Th & $\alpha$+$^{224}$Ra & 5.520  & 0  & $2.78\times 10^{-29}$ & 4.1 & 5.3 \\
        & $^{20}$O+$^{208}$Pb & 44.723 & 0 & $5.56\times10^{-35}$ & 4.0 & 10.2 \\
        
        &&&&&& \\
        $^{231}$Pa & $\alpha$+$^{227}$Ac & 5.150  & 0  & $5.55\times 10^{-32}$ & 4.2 & 8.1 \\
        & $^{23}$F+$^{208}$Pb & 51.888 & 1 & $1.95\times10^{-36}$ & 6.8 & 10.9 \\
        
        &&&&&& \\
        $^{232}$U & $\alpha$+$^{228}$Th & 5.413 & 0  & $6.60\times10^{-31}$ & 4.2 & 8.1 \\
        & $^{24}$Ne+$^{208}$Pb & 62.310 & 0 & $2.78\times10^{-32}$ & 3.9 & 10.9 \\
        
        &&&&&& \\
        $^{233}$U & $\alpha$+$^{229}$Th & 4.908 & 0 & $4.13\times10^{-34}$ & 4.1 & 8.4 \\
        & $^{24}$Ne+$^{209}$Pb & 60.485 & 2 & $5.41\times10^{-35}$ & 11.0 & 11.7 \\
        & $^{28}$Mg+$^{205}$Hg & 74.226 & 3 & $7.00\times10^{-36}$ & 15.7 & 13.1 \\
        
        &&&&&& \\
        $^{235}$U & $\alpha$+$^{231}$Th & 4.678  & 1  & $7.94\times 10^{-36}$ & 11.8 & 8.4 \\
        & $^{25}$Ne+$^{210}$Pb & 57.683 & 3 & $8.57\times10^{-40}$ & 15.6 & 12.5 \\
        & $^{28}$Mg+$^{207}$Hg & 72.425 & 1 & $3.65\times10^{-38}$ & 7.2 & 13.8 \\
        
        &&&&&& \\
        $^{237}$Np & $\alpha$+$^{233}$Pa & 4.958 & 1 & $2.19\times10^{-34}$ & 11.8 & 8.0 \\
        & $^{30}$Mg+$^{207}$Tl & 74.790 & 2 & $1.34\times10^{-36}$ & 10.5 & 12.3 \\
        
        &&&&&& \\
        $^{236}$Pu & $\alpha$+$^{232}$U & 5.867 & 0  & $2.41\times10^{-29}$ & 4.0 & 8.1 \\
        & $^{28}$Mg+$^{208}$Pb & 79.669 & 0 & $3.86\times10^{-31}$ & 3.8 & 11.6 \\
        
        &&&&&& \\
        $^{241}$Am & $\alpha$+$^{237}$Np & 5.637  & 1  & $3.75\times 10^{-31}$ & 11.9 & 8.2 \\
        & $^{34}$Si+$^{207}$Tl & 93.926 & 3 & $1.00\times10^{-32}$ & 14.4 & 12.0 \\
        
        &&&&&& \\
        $^{242}$Cm & $\alpha$+$^{238}$Pu & 6.215  & 0  & $1.88\times10^{-28}$ & 4.1 & 8.1 \\
        & $^{34}$Si+$^{208}$Pb & 96.510 & 0 & $4.59\times10^{-31}$ & 3.6 & 11.5 \\
        \hline
    \end{tabular}
    \caption{Comparison of the penetration probabilities calculated within the Mumbai ($M$) and Chaudhury ($Ch$) approach. The effect of nonlocality is expressed in terms of a percentage increase in the penetration factor within the $M$ or $Ch$ model as compared to the double folding model and is given by $PC^{(NL)}=\big(P^{(NL)}-P^{(DF)}\big)\times 100/P^{(DF)}$.}
    \label{tab:penetration_prob}
\end{table*}
In the section discussing the different approaches for nonlocality, we mentioned one of the earliest works \cite{chaudhuryPRC,chaudhuryPRL}  
studying the effects of the nonlocal strong interaction in $\alpha$ decay. The author evaluated the penetration proba\-bilities for the $\alpha$ decay of several nuclei using Eq. (\ref{eq:ChauP1}). In Table \ref{tab:penetration_prob}, we compare the penetration factors obtained in the Mumbai approach studied in this work with that of Ref. \cite{chaudhuryPRC}. However, in order to perform a consistent comparison we replace the Igo potential used in Ref. \cite{chaudhuryPRC} by the DF one used here and similarly the point-like Coulomb interaction in Eq. (\ref{eq:ChauP1}) is replaced by that obtained using the DF procedure too. Apart from this, recalling that applying the JWKB procedure to the radial equation leads to an improper behavior of the solution near the origin unless one replaces the centrifugal term by the Langer modified term mentioned earlier \cite{Langer1937}, we replace $l(l+1)$ in Eq. (\ref{eq:ChauP1}) by $(l + 1/2)^2$. The Bohr-Sommerfeld quantization, which was not considered in Ref. \cite{chaudhuryPRC}, is also taken into account to obtain a modified version of Eq. (\ref{eq:ChauP1}) given by
\begin{multline}
    P^{\rm Ch} = \exp\biggl \{ -2 \sqrt{2\mu \over \hbar^2} \int_{r_2}^{r_3} \biggl [ \Big(\lambda\, V_{N}(r)\\+V_{C}(r)-E\Big)\epsilon(r)+\frac{\hbar^{2}(l+1/2)^2}{2\mu r^{2}} \biggr ]^{1/2} {\rm d}r \biggr \}
\end{multline}
where $V_N$ is the attractive strong potential whose depth is modified by the factor $\lambda$ due to the Bohr-Sommerfeld condition. The turning points $r_2$ and $r_3$ are given by $V(r) = E$ as before with $E$ taken to be the $Q$ value of the decay. 

The penetration factors in both models are found to increase due to the inclusion of nonlocality in the nuclear potential. This is consistent with the fact that the half-lives in general decrease due to nonlocality. The effect is expressed in terms of a percentage change in the penetration probability given by $PC^{(NL)}=\big(P^{(NL)}-P^{(DF)}\big)\times 100/P^{(DF)}$. The order of magnitude of the changes are similar in both the models considered. However, the results within the Mumbai approach are sensitive to the $l$ value in the decay. This is expected since the potential defined by Eq. (\ref{eq:mumbaipot}), obtained within the Mumbai approach is $l$ dependent.  

\section{Summary}\label{summar}
In an earlier work \cite{Johan2019}, the effect of including nonlocality in the nuclear interaction potential was found to decrease the $s$-wave ($l$ = 0) $\alpha$ decay half-lives of spherical nuclei. Results in Ref. \cite{Johan2019} indicated that whereas the energy-dependent Perey-Buck (PB) model produced a decrease of 20--40 \% in the half-lives, $t_{1/2}$, the energy-independent Mumbai (M) potential decreased the half-lives by only 2--4 \%. Since both the Perey-Buck and Mumbai potentials reproduce the scattering data quite well, this large difference in the decrease of $t_{1/2}$ was surprising. The Mumbai potential is $l$ dependent whereas the Perey-Buck potential does not depend on $l$. In the calculation of cross sections, where one usually sums over all possible $l$ values to compare with data, the differences between the Mumbai and Perey-Buck potentials do not become quite evident. However, if one considers a nuclear decay, one is actually picking up a particular value of $l$. Hence, in the present work, different nuclear decays involving non-zero values of $l$ were investigated. Considering nuclei with $l \ne$ 0 to be spherical is not always a good approximation. Hence, the above studies were carried out by using a deformed potential whenever necessary. The latter allowed us to also study the effect of deformation on nonlocality.

When a nucleus has the possibility of decaying by emitting an $\alpha$ particle ($^4$He nucleus) or a light cluster (such as $^{14}$C, $^{24}$Ne, $^{28}$Mg, etc.), one can consider the decaying nucleus to be a preformed cluster of an $\alpha$ and a heavy daughter or a light nucleus and a not-so-heavy daughter ($^{232}$U considered as a preformed cluster of $^4$He + $^{228}$Th or $^{24}$Ne + $^{208}$Pb, for example). Would the effects of nonlocality (and deformation) show up differently in the two scenarios? To answer this question, the half-lives were calculated for those nuclei which can decay by emitting both an $\alpha$ or a light nucleus. The exercise was rewarding not only in the sense of finding the different manifestation of nonlocality in the two kinds of decays of the same parent nucleus but also in noting the differences in the two nonlocality models studied here.

Since the detailed observations are many, we summarize here the global conclusions drawn from the results: 
\renewcommand{\labelenumi}{(\roman{enumi})}
\begin{enumerate}
    \item In the tunneling decay of a nucleus which is considered to be a preformed cluster of the tunneling light nucleus and a heavy daughter, it is important to consider the nonlocality in the interaction potential between the two nuclei forming the cluster. 
    \item Including the effects of deformation of the light nucleus in cluster decay or the heavy daughter in $\alpha$ decay, both lead to a decrease in the half-lives of the nuclei. The amount of decrease is decided by the magnitude of the deformation parameters but not affected by the sign of the parameters. 
    \item Sensitivity of the results to the $l$ value (angular momentum) in the decay is manifested within the Mumbai model for nonlocality. 
    \item The combined effect of including deformation and nonlocality (on the half-lives of nuclei) is in ge\-neral quite large within both models of nonlocality stu\-died here. 
    \item Phenomenologically determined cluster preformation factors display a linear dependence on the $Q$-value or the energy carried by the emitted $\alpha$ or light cluster.   
\end{enumerate}

Finally, we add that given the nature of the present work with the many ingredients entering into the calculations of half-lives, there is surely scope for better calculations. For example, one could improve the tunneling model by including the intrinsic degrees of freedom of the composite tunneling object as shown in Ref. \cite{Bertulani2007}. Though the nonlocality in the two body interaction has been taken into account, there remains the possibility of including the nonlocality inside the interacting nuclei. Finally, one could replace the semiclassical approach used here by a fully quantum mechanical one or study the nuclear decay within different approaches as in Ref. \cite{Ni-Ren2015}. Future investigations in this direction can prove useful for a better understanding of nuclear structure as well as nonlocal nuclear interactions. 

\begin{acknowledgments}
D.F.R-G. thanks the Faculty of Science, Universidad de los Andes, Colombia, for financial support through Grant No. INV-2020-103-2053. N.J.U acknowledges financial support from SERB, Government of India (Grant No. YSS/2015/000900).
\end{acknowledgments}



%

\end{document}